\documentclass[preprint]{aastex631}

\newcommand \versnum {4}  
\newcommand{\fwidth}{8.0cm}   
\newcommand{\fwidthb}{5.8cm}   
\newcommand{\fwidthc}{8.0cm}   
\usepackage{color}
\usepackage{natbib}
\usepackage{amsmath}
\usepackage{nicefrac}
\tightenlines

\newcommand \bahat      {\boldsymbol{\hat{a}}}

\newcommand \bD         {\boldsymbol{D}}
\newcommand \bE         {\boldsymbol{E}}

\newcommand \bkhat      {\hat{\bf k}}

\newcommand \bP         {\boldsymbol{P}}

\newcommand \beq        {\begin{equation}}
\newcommand \beqa	{\begin{eqnarray}}

\newcommand \cm         {\,{\rm cm}}

\newcommand \eeq	{\end{equation}}
\newcommand \eeqa	{\end{eqnarray}}

\newcommand \ext        {{\rm ext}}

\newcommand \gtsim	{\gtrsim}		 

\newcommand \lambdap    {\lambda_{\rm p}}

\newcommand \ltsim	{\lesssim}		 

\newcommand \Ndip       {N_{\rm dip}}              

\newcommand \pfirmax    {[p_{\rm em}({\rm FIR})]_{\rm max}}

\newcommand \poromacro  {{\cal P}_{\rm macro}}
\newcommand \poromicro  {{\cal P}_{\rm micro}}

\newcommand \Asymm      {{\cal A}}
\newcommand \Stretch    {{\cal S}}

\newcommand \sigmap     {\sigma_{\rm p}}

\newcommand \aeff       {a_{\rm eff}}
\newcommand \aeffp      {a_{\rm eff,p}}

\newcommand \Cabs       {C_{\rm abs}}
\newcommand \Cext       {C_{\rm ext}}

\newcommand \CextPSA    {C_{\rm ext,PSA}}

\newcommand \Csca       {C_{\rm sca}}

\newcommand \extran     {{\rm ext,ran}}

\newcommand \extPSA     {{\rm ext,PSA}}

\newcommand \polPSA     {{\rm pol,PSA}}
\newcommand \falign     {f_{\rm align}}
\newcommand \DDA        {{\rm DDA}}
\newcommand \EMT        {{\rm EMT}}
\newcommand \PSA        {{\rm PSA}}

\newcommand \solid      {{\rm sol}}
\newcommand \thetas     {\theta_{\rm s}}

\newcommand \sgn        {{\rm sgn}}
\newcommand \fvac       {f_{\rm vac}}
\newcommand \epssol     {\epsilon_{\rm sol}}
\newcommand \Brug       {{\rm Brug}}
\newcommand \LLL        {{\rm LLL}}

\newcommand{\omittext}[1]{}

\countdef\decade=200
\advance\decade by \year
\countdef\hours=201
\advance\hours by \time
\divide\hours by 60
\countdef\mins=202
\advance\mins by \hours
\multiply\mins by 60
\multiply\hours by 100
\countdef\miltime=203
\advance\miltime by \hours
\advance\miltime by \time
\advance\miltime by -\mins
\renewcommand\today{\number\decade.\number\month.\number\day.\number\miltime}
\begin{document}

\title{%
        \vspace*{-2.0em}
        {\normalsize\rm Published in {\it The Astrophysical Journal} {\bf985}:10 (2025 May 20)}\\
        \vspace*{1.0em}
        {\bf The Spheroidal Analog Method for
         Modeling Irregular Porous Aggregates}
	}

\author[0000-0002-0846-936X]{B.~T.~Draine}
\affiliation{Dept.\ of Astrophysical Sciences,
  Princeton University, Princeton, NJ 08544, USA}
\affiliation{Institute for Advanced Study,
  Princeton, NJ 08540, USA}

\email{draine@astro.princeton.edu}

\begin{abstract}
It is shown that the optical properties of an irregular porous grain
with effective radius $\aeff\ltsim 3\lambda$ (where $\lambda$ is the
wavelength) can be well approximated by a ``spheroidal analog'': a
spheroid with appropriate axial ratio and size, with a dielectric
function obtained from an effective medium theory.  Prescriptions for
specifying the axial ratio and porosity of the spheroidal analog,
based on simple geometric properties, are given.  The accuracy of the
spheroidal analog method is studied for irregular grains with a
range of structures and porosities.  Different effective medium
theories are compared; Bruggeman's theory is found to give the best
results.  The accuracy of the spheroidal analog method justifies the
use of spheroids for modeling absorption, scattering and polarization
by interstellar, circumstellar, or interplanetary dust.
\end{abstract}
\keywords{
          interstellar dust (836),
          radiative transfer (1335)}

\section{Introduction
         \label{sec:intro}}

Dust grains play a major role in the chemistry and dynamics of the
interstellar medium (ISM).  In diffuse regions, dust dominates the
formation of molecular hydrogen, provides shielding of molecules from
far-ultraviolet radiation, heats the diffuse ISM via photoelectric
emission, and couples radiation pressure to the gas.  In regions with
low fractional ionization, dust grains affect the level of ionization
by trapping ions and electrons, and by providing a pathway for
non-radiative recombination. In dense molecular regions with low
fractional ionization, charged dust grains can be important for
coupling magnetic fields to the neutral gas.  In addition,
observations of polarized extinction and infrared emission provide
valuable diagnostics of physical conditions, including the magnetic
field.

Despite the recognized importance of dust, the physical structure of
interstellar grains remains uncertain.  The observed polarization of
starlight requires that the grains be significantly aspherical.  Some
authors have approximated grains by simple shapes such as spheroids
\citep[e.g.,][]{Kim+Martin_1995b,
Voshchinnikov+Das_2008,
Draine+Fraisse_2009,
Das+Voshchinnikov+Ilin_2010,
Siebenmorgen+Voshchinnikov+Bagnulo_2014,
Siebenmorgen+Voshchinnikov+Bagnulo+Cox_2017,
Hensley+Draine_2023},
and models for interstellar dust employing homogeneous spheroids
\citep{Hensley+Draine_2023} are compatible with current observational
constraints.

However, other authors
\citep[e.g.,][]{Abadi+Wickramasinghe_1976, Jones_1988,
  Mathis+Whiffen_1989, Ossenkopf_1993, Kimura+Mann+Wehry_1999}
have argued that on-going coagulation in interstellar clouds should
result in highly porous grains with very irregular structures.
Laboratory studies of interplanetary dust particles (IDPs) collected
in the stratosphere show that some -- the anhydrous chondritic IDPs --
have highly porous, aggregate structures \citep[see,
  e.g.,][]{Bradley_2003b}.  Although these IDPs are much larger than
representative interstellar grains, similar coagulation processes in
the ISM may assemble nanoparticles into submicron aggregates.

\begin{figure}
\newcommand{\figheight}{5.0cm}   
\newcommand{\triml}{0.7cm}
\newcommand{\trimr}{0.0cm}
\newcommand{\trimb}{8.0cm}
\newcommand{\trimt}{2.5cm}
\newcommand{\negspace}{-0.3cm}
\newcommand{\midspace}{0.3cm}
\begin{center}
\includegraphics[angle=0,height=\figheight,
                 clip=true,trim={\triml} {\trimb} {\trimr} \trimt]
{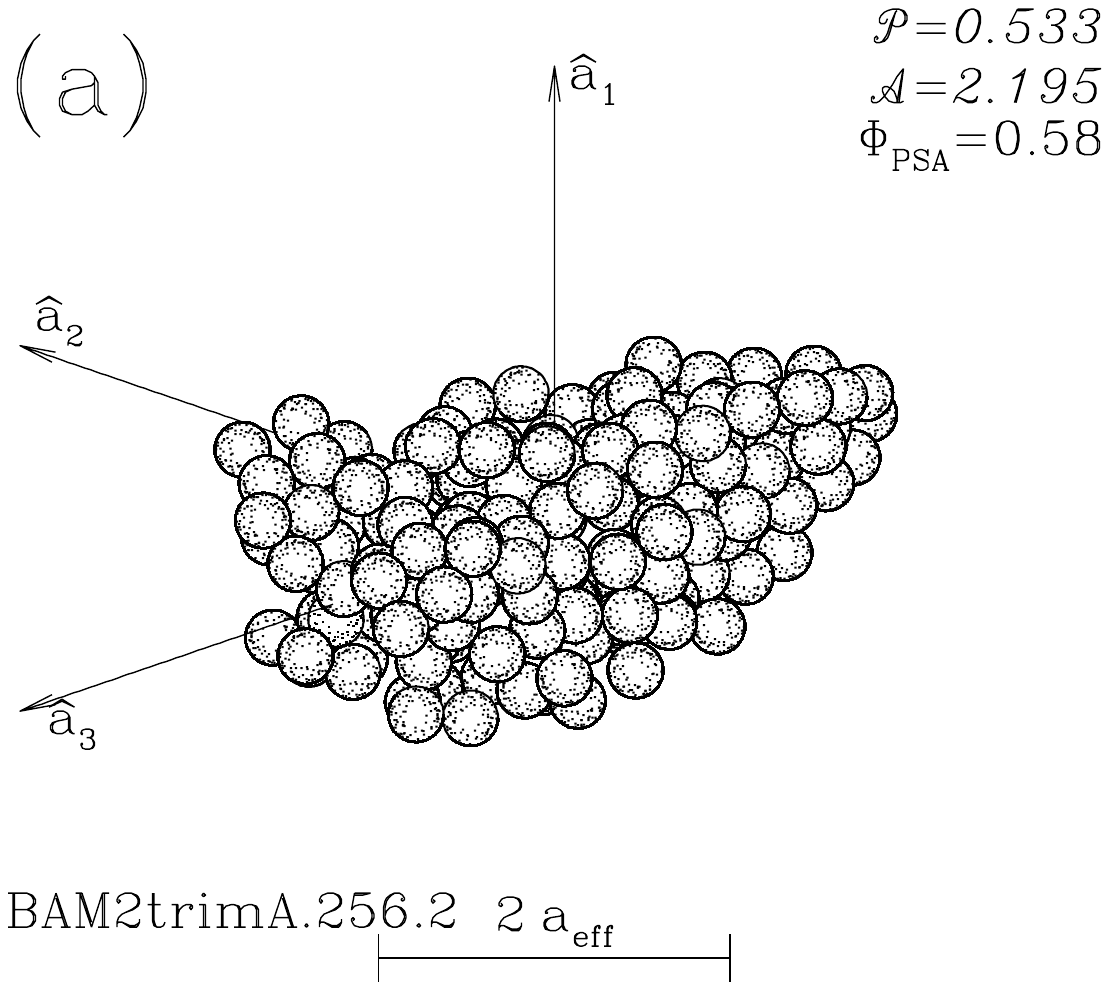}   
\hspace*{\negspace}
\includegraphics[angle=0,height=\figheight,
                 clip=true,trim={\triml} {\trimb} {\trimr} \trimt]
{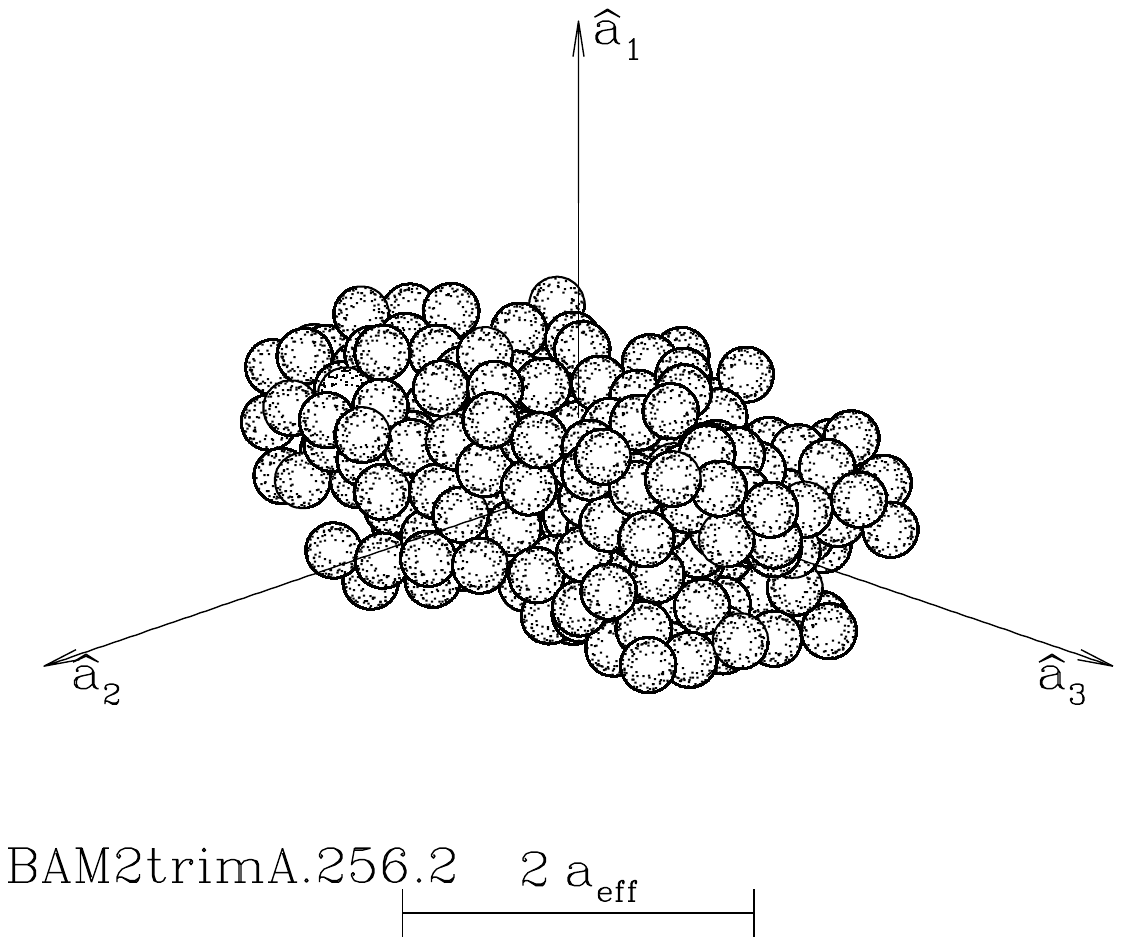}   
\hspace*{\negspace}
\includegraphics[angle=0,height=\figheight,
                 clip=true,trim={\triml} {\trimb} {\trimr} \trimt]
{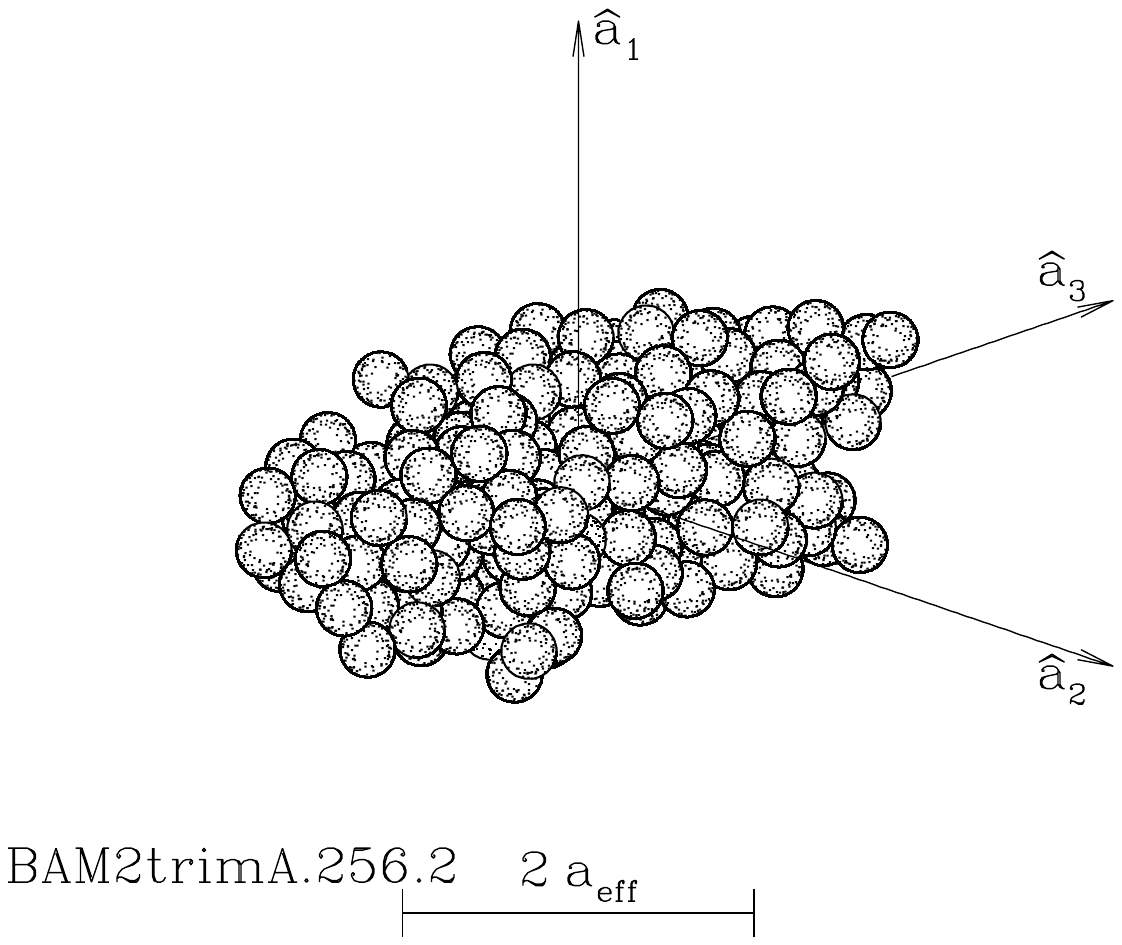}   

\vspace*{-1.0cm}

\includegraphics[angle=0,height=\figheight,
                 clip=true,trim={\triml} {\trimb} {\trimr} \trimt]
{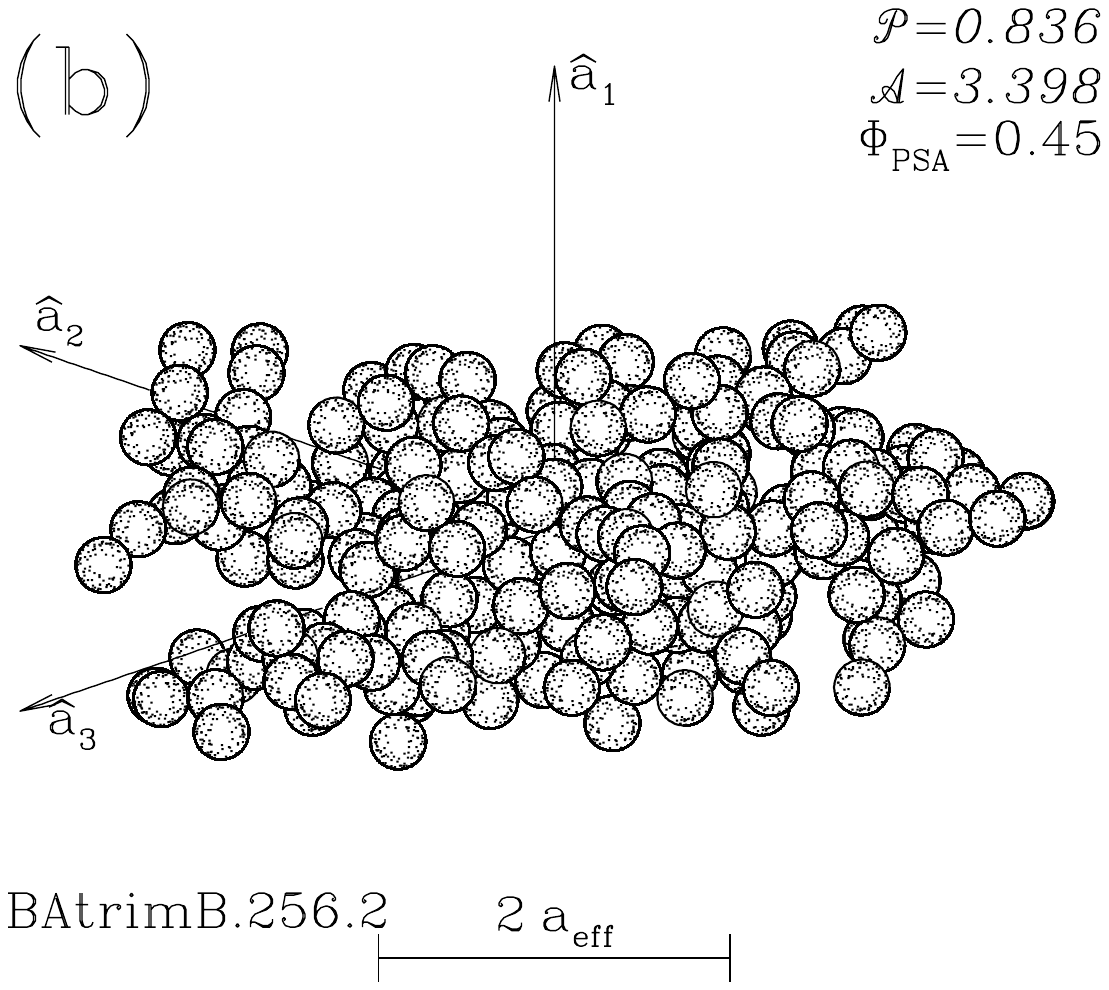}   
\hspace*{\negspace}
\includegraphics[angle=0,height=\figheight,
                 clip=true,trim={\triml} {\trimb} {\trimr} \trimt]
{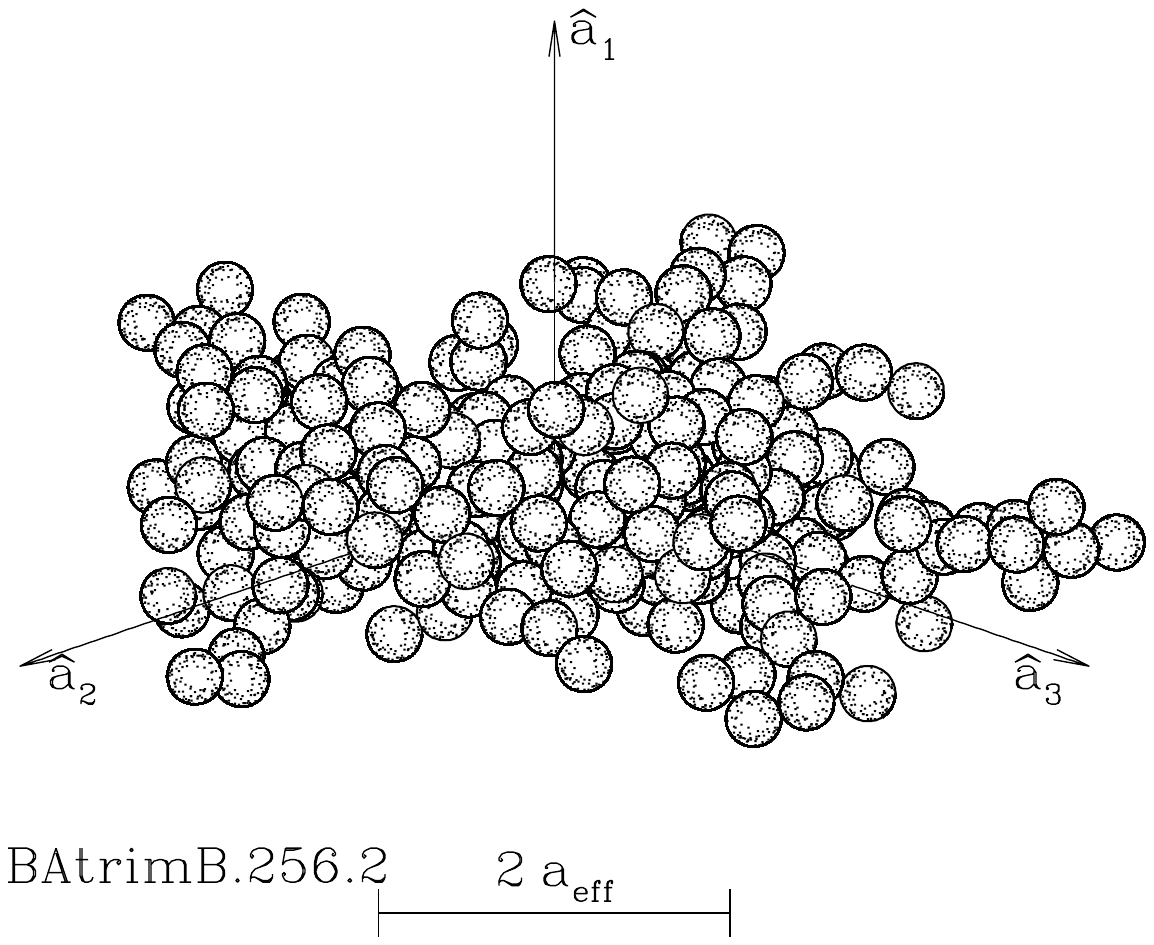}   
\hspace*{\negspace}
\includegraphics[angle=0,height=\figheight,
                 clip=true,trim={\triml} {\trimb} {\trimr} \trimt]
{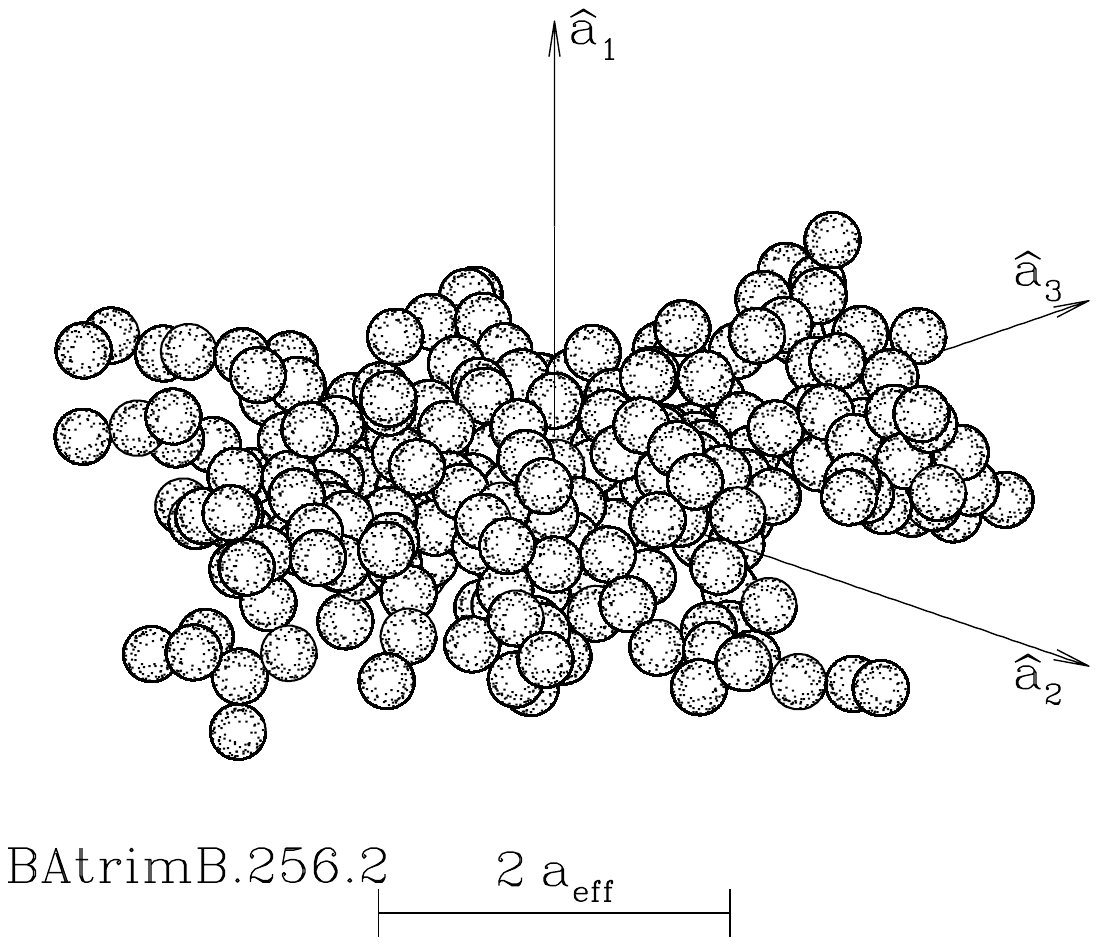}   

\caption{\label{fig:bam1trimB}\footnotesize Two examples of irregular
  porous aggregates studied by \citet{Draine_2024b}: {\bf (a)} case 1
  aggregate (see Table \ref{tab:targets}) with $\poromacro=0.533$,
  {\bf (b)} case 6 aggregate with $\poromacro=0.836$.  For each
  aggregate, 3 views are shown.  $\bahat_1$ is the principal axis of
  largest moment of inertia.
    }
\end{center}
\end{figure}

If interstellar grains are indeed irregular aggregates, we need
methods for modeling their optical properties.  Unfortunately, direct
calculation of absorption and scattering by irregular aggregates (such
as those in Figure \ref{fig:bam1trimB}) requires computational methods
such as the discrete dipole approximation (DDA) to solve Maxwell's
equations for incident plane waves \citep{Draine+Flatau_1994}.  Such
calculations can be computationally challenging, limiting our ability
to study models that include such porous grains.  Faster methods, even
if approximate, are needed.

Furthermore, the exact shapes of such aggregates are unknown (and
unknowable), and the possibilities are infinite.  An approach is
needed that can focus on important structural parameters rather than
specific irregular structures.

In the diffuse ISM, grains have overall sizes $D\ltsim 1\micron$, and
any porous substructure within the grain will necessarily be on
smaller (submicron) scales.  For optical and infrared radiation with
wavelength $\lambda\gtsim 0.5\micron$, one might hope to approximate
the actual porous grain material by a homogeneous medium characterized
by a suitable ``effective'' dielectric function $\epsilon_{\rm EMT}$
obtained from an ``effective medium theory'' (EMT).

\citet{Voshchinnikov+Videen+Henning_2007} studied the extinction and
scattering properties of porous spheres and randomly-oriented porous
spheroids, and tested different EMTs to approximate the porous
material; the EMT due to \citet{Bruggeman_1935} was found to give the
best results.

\citet{Shen+Draine+Johnson_2008,Shen+Draine+Johnson_2009} calculated
extinction and scattering by various random aggregates.  They found
that extinction and scattering by randomly-oriented aggregates can be
approximated by homogeneous spheres with suitably chosen total volume
and a dielectric function obtained using the Bruggeman EMT.
\citet{Lagarrigue+Jacquier+Debayle+etal_2012} similarly studied
scattering by a varied set of aggregates, using the
\citet{Maxwell_Garnett_1904} EMT and spheres with volume equal to that
of a ``convex hull'' circumscribing the aggregate.

Here we develop and test a new approach -- the ``spheroidal analog
method'' (SAM) -- for approximating electromagnetic scattering and
absorption by irregular aggregates.  We go beyond previous work by
studying the polarization properties that arise when the aggregates
are not randomly oriented.  We show how spheroidal shapes with
appropriate axial ratios and porosities can be used to approximate
both the total extinction and the polarization properties of irregular
aggregates.  The accuracy of the approximation is tested by comparing
to ``exact'' results for the aggregates calculated with the DDA.

The four most commonly used EMTs are summarized in Section
\ref{sec:emt}.  In Section \ref{sec:Testing EMTs} we test the accuracy
of approximating irregular porous aggregates by ``spheroidal
analogs", with specified shape and size, using the different EMTs.
We show that the spheroidal analog method (SAM) using the Bruggeman
EMT can usefully approximate the optical properties of irregular
porous aggregates, for wavelengths from the vacuum ultraviolet
($0.1\micron$) to the far-infrared ($100\micron$).  Our results are
discussed in Section \ref{sec:discuss}, and summarized in Section
\ref{sec:summary}.

\section{\label{sec:emt} Effective Medium Theories}

When a medium is inhomogeneous only on scales small compared to the
wavelength $\lambda$, one may characterize its volume-averaged
response to an applied oscillating electric field by an effective
dielectric function $\epsilon_{\rm EMT}(\lambda)$ determined by the
dielectric functions and volume filling factors of the constituents.
``Effective medium theory'' (EMT) refers to the relation between
$\epsilon_{\rm EMT}$ and the actual dielectric functions of the
constituent materials.

\citet{Bohren+Huffman_1983} have a brief and lucid discussion of EMTs;
see \citet{Sihvola_1999} or \citet{Chylek+Videen+Geldart+etal_2000}
for more thorough reviews.  For monochromatic radiation, the effective
(complex) dielectric function
$\epsilon_{\rm EMT}(\lambda)$
is such that
\beq
\langle \bD\rangle \equiv \langle \bE \rangle + 4\pi\langle\bP\rangle
=
\epsilon_{\rm EMT}\langle\bE\rangle 
~~~, 
\eeq
where $\langle \bE\rangle$ and $\langle\bP\rangle$ are the
volume-averaged electric field and polarization; the averaging is over
regions small compared to $\lambda$ but large compared to the
characteristic scale of the inhomogeneities.  Here we limit
consideration to two constituents, referring to one as ``matrix'', and
the other as ``inclusions''.  Let ``matrix'' and ``inclusions'' be
characterized by dielectric functions $\epsilon_{\rm mat}$ and
$\epsilon_{\rm inc}$, with volume filling factors $f_{\rm mat}$ and
$f_{\rm inc}=1-f_{\rm mat}$, respectively.

\subsection{Maxwell Garnett Theory: MG1 and MG2}

The approach due to \citet{Maxwell_Garnett_1904,Maxwell_Garnett_1906}
treats matrix and inclusions asymmetrically.  If the inclusions are
taken to be spherical, the effective dielectric function
$\epsilon_{\rm MG}$ is the solution to
\beq
0 = \left(\frac{\epsilon_{\rm mat}-\epsilon_{\rm MG}}
                    {\epsilon_{\rm MG}+2\epsilon_{\rm mat}}\right)
+ f_{\rm inc}\left(\frac{\epsilon_{\rm inc}-\epsilon_{\rm mat}}
                        {\epsilon_{\rm inc}+2\epsilon_{\rm mat}}\right)
~~~.
\eeq
For a porous material consisting of solid and vacuum, the usual
approach is to take the solid regions to be the ``matrix'', and the
vacuum to be the ``inclusions''; we refer to this as ``MG1'':
\beq
\epsilon_{\rm MG1} = 
\frac{\epssol[2\epssol+1-2\fvac(\epssol-1)]}
     {2\epssol+1+\fvac(\epssol-1)}
~~~,
\eeq
where $\epssol$ is the dielectric function of the solid material, and
$\fvac$ is the vacuum filling factor.

Alternatively, one may choose to treat the vacuum as the ``matrix'',
and the solid material as the ``inclusions''; we refer to this as
MG2:\footnote{\citet{Voshchinnikov+Videen+Henning_2007} refer to this
as ``inverse Garnett''.}
\beq
\epsilon_{\rm MG2} = 
\frac{\epssol(3-2\fvac)+2\fvac}
     {3+\fvac(\epssol-1)}
~~~.
\eeq
The MG2 approach might seem reasonable for very high porosity media
with $\fvac>0.5$, where most of the volume is vacuum.

\subsection{Bruggeman Theory}

The EMT due to \citet{Bruggeman_1935} treats the two components
symmetrically: the effective dielectric function $\epsilon_\Brug$
is the solution to
\beq \label{eq:Brugg}
0 = (1-f_{\rm inc}) 
    \left(\frac{\epsilon_{\rm mat}-\epsilon_\Brug}
               {\epsilon_{\rm mat}+2\epsilon_\Brug}\right)
     + f_{\rm inc}
    \left(\frac{\epsilon_{\rm inc}-\epsilon_\Brug}
               {\epsilon_{\rm inc}+2\epsilon_\Brug}\right)
~~~.
\eeq
Thus, setting $f_{\rm inc}=\fvac$, $\epsilon_{\rm inc}=1$, and
$\epsilon_{\rm mat}=\epssol$:
\beqa \label{eq:epsilon_Brug}
\epsilon_\Brug &~~=~~& \frac{B+(B^2+2\epssol)^{1/2}}{2}
\\ \label{eq:B_Brug}
B&\equiv& \frac{\epssol(2-3\fvac)+3\fvac-1}{2}
~~~. 
\eeqa
%

\begin{figure}
\begin{center}
\includegraphics[angle=0,width=\fwidthc,
                 clip=true,trim=0.5cm 0.5cm 0.0cm 0.5cm]
{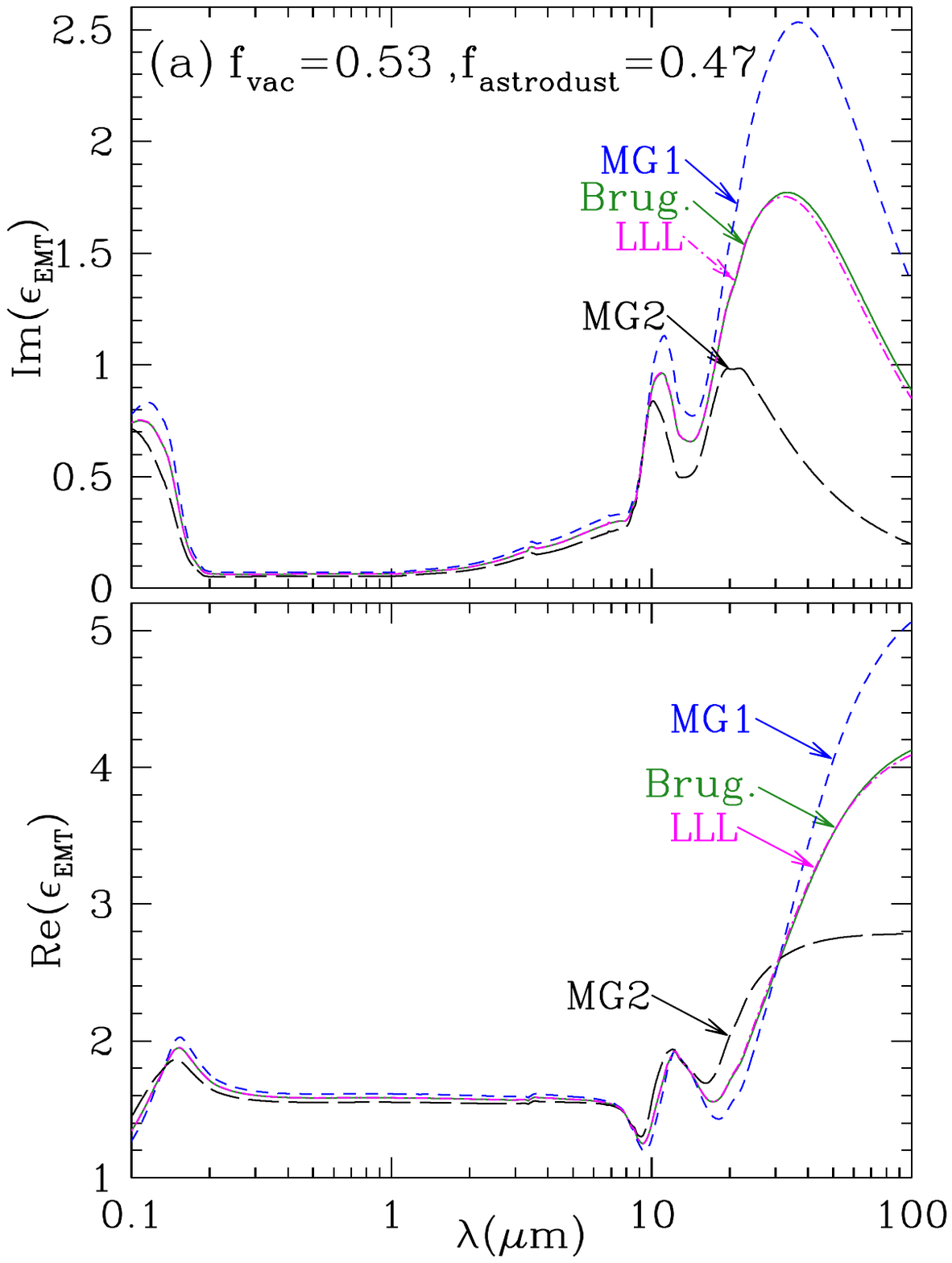}
\includegraphics[angle=0,width=\fwidthc,
                 clip=true,trim=0.5cm 0.5cm 0.0cm 0.5cm]
{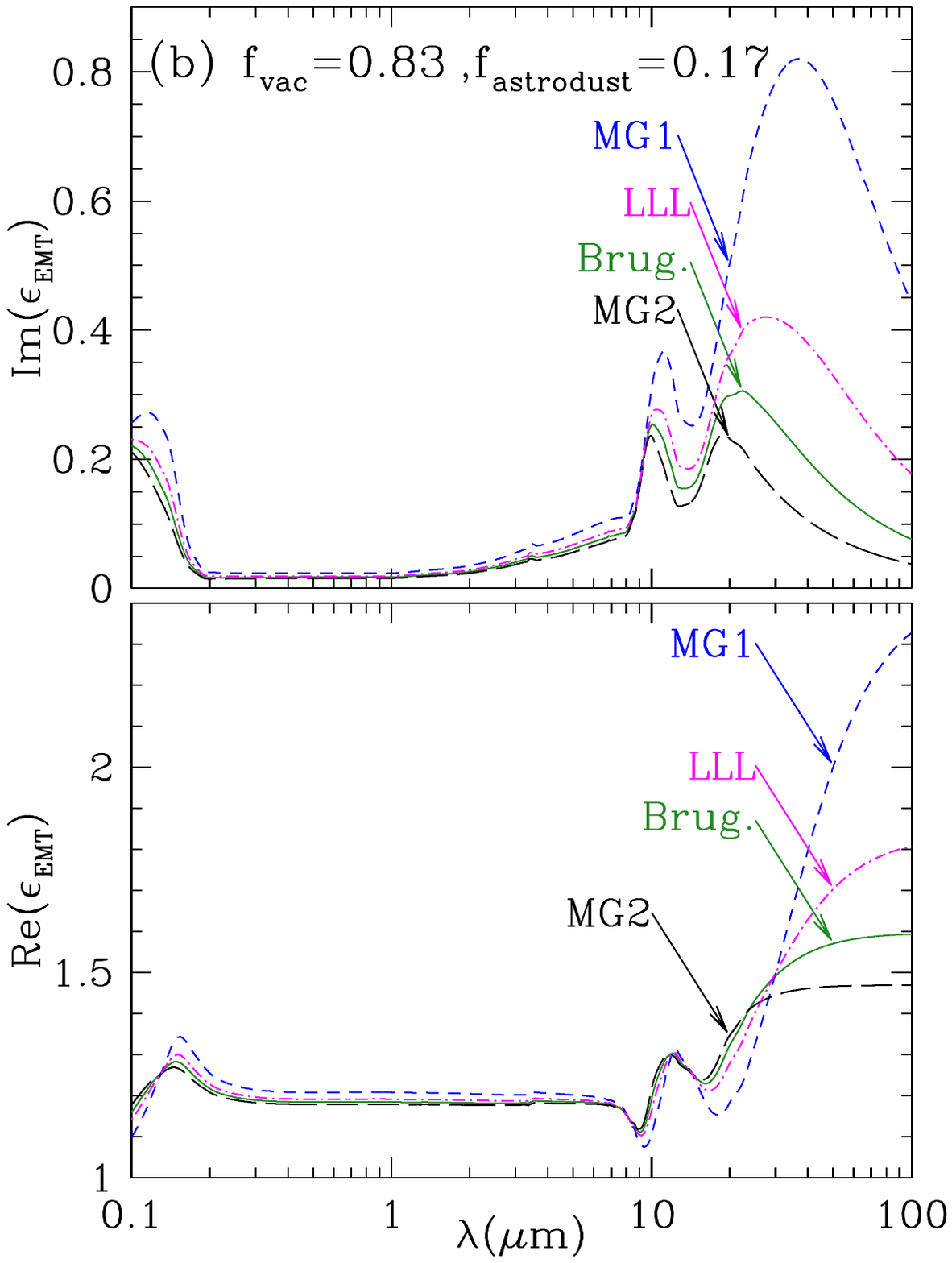}
\caption{\label{fig:eps_emt}\footnotesize Effective dielectric
  functions for porous material consisting of ``astrodust'' solid plus
  vacuum, for the Bruggeman (Brug.), Landau-Lifshitz-Looyenga (LLL),
  and Maxwell Garnett EMTs (MG1 and MG2; see text).  Results are shown
  for mixtures of {\bf (a)} 47\% astrodust + 53\% vacuum, and {\bf
    (b)} 17\% astrodust + 83\% vacuum.
  }
\end{center}
\end{figure}
\subsection{Landau-Lifshitz-Looyenga Theory}

\citet{Landau+Lifshitz_1960} and \citet{Looyenga_1965} proposed a
simple mixing rule that treats the matrix and inclusions
symmetrically:
\beq \label{eq:LLL}
\epsilon_\LLL = 
\left[f_{\rm mat}\epsilon_{\rm mat}^{1/3} + 
      f_{\rm inc}\epsilon_{\rm inc}^{1/3}\right]^3
~~~.
\eeq
Studies of the optical properties of wafers of compressed SiO$_2$
nanopowder \citep{Nzie+Blanchard+Genevois+DeSousaMeneses_2019} found
that $\epsilon_{\rm LLL}$ is in better agreement with measured
infrared reflectivities than the Bruggeman or Maxwell Garnett models.

Figure \ref{fig:eps_emt} shows $\epsilon_{\rm Brug}(\lambda),
\epsilon_{\rm MG1}(\lambda)$, $\epsilon_{\rm MG2}(\lambda)$, and
$\epsilon_{\rm LLL}(\lambda)$ for $\fvac=0.53$ and $0.83$ (the
smallest and largest values of $\fvac$ studied below), for
$\epsilon_\solid(\lambda)=\epsilon_{\rm Ad}(\lambda)$, where
$\epsilon_{\rm Ad}(\lambda)$ is the dielectric function estimated for
``astrodust'' material
\citep{Draine+Hensley_2021a}.\footnote{%
We take $\epsilon_{\rm AD}(\lambda)$ to be the ``astrodust''
dielectric function for microporosity 0.2 and 1.4:1 oblate spheroids,
obtained by \citet{Draine+Hensley_2021a}.  $\epsilon_{\rm
  Ad}(\lambda)$ is available at
\url{https://doi.org/10.34770/9ypp-dv78}.}
For $\lambda<10\micron$, $\epsilon_{\rm MG1}$, $\epsilon_{\rm MG2}$,
$\epsilon_\Brug$, and $\epsilon_\LLL$ are similar.  However, at
$\lambda\gtsim10\micron$, where $|\epsilon_{\rm Ad}(\lambda)|$ becomes
large, the $\epsilon_{\rm EMT}(\lambda)$ can differ appreciably,
leading to significant differences in calculated cross sections.
$\epsilon_\Brug$ and $\epsilon_\LLL$ are intermediate between
$\epsilon_{\rm MG1}$ and $\epsilon_{\rm MG2}$.

\subsection{Other Effective Medium Theories}

While the EMTs due to Maxwell Garnett and Bruggeman are best known,
others have been proposed \citep[see reviews by][]{Sihvola_1999,
  Chylek+Videen+Geldart+etal_2000}.  In addition to the Bruggeman,
Maxwell Garnett, and Landau-Lifshitz-Looyenga EMTs,
\citet{Voshchinnikov+Videen+Henning_2007} tested the EMTs due to
\citet{Lichtenecker_1926} and
\citet{Birchak+Gardner+Hipp+Victor_1974}, but found Bruggeman's EMT to
give the best results for porous spheres and randomly-oriented
spheroids.  \citet{Bohren_1986} and
\citet{Mishchenko+Dlugach+Liu_2016} discusssed the applicability of
EMTs when the inhomogeneities have sizes that are not much smaller
than $\lambda$.

The present study is limited to the Maxwell Garnett, Bruggeman, and
Landau-Lifschitz-Looyenga EMTs.

\section{\label{sec:Testing EMTs}
              The Spheroidal Analog Method (SAM)}

Can electromagnetic scattering and absorption by an irregular, porous
structure be calculated by approximating it as a homogeneous spheroid
with a suitable size and axis ratio, and an effective dielectric
constant?

\subsection{Targets: Irregular Grains}

\citet{Draine_2024b} used the DDA to calculate scattering and
absorption by a number of irregular aggregates of equal-size solid
spheres.  Random aggregates were generated using different aggregation
schemes (``BA'', ``BAM1'', and ``BAM2''); some of these were then
``trimmed'' by two different prescriptions (``trimA'' and ``trimB'')
to make them more asymmetric \citep[see][]{Draine_2024b}.  We will use
such aggregates here, each composed of 256 equal-size solid spheres.
The material in the spheres has density $\rho_\solid$.

The dielectric function $\epsilon_\solid(\lambda)$ for the solid
material in the constituent spheres was taken to be the ``astrodust''
dielectric function $\epsilon_{\rm Ad}(\lambda)$ \citep[for $b/a=1.4$
  and $\poromicro=0.2$;][]{Draine+Hensley_2021a}.

An aggregate of mass $M$ has solid volume $V_\solid = M/\rho_\solid$,
where $\rho_\solid$ is the density of the solid material.  The
``effective radius'' of the aggregate is defined by the radius of an
equal-volume sphere:
\beq \label{eq:aeff}
\aeff\equiv\left(\frac{3V_\solid}{4\pi}\right)^{1/3}
~~~.
\eeq
As discussed by \citet{Draine_2024b}, the shape of an aggregate can be
characterized by three geometric parameters -- $\poromacro$, $\Asymm$,
and $\Stretch$ -- obtained from the eigenvalues of the moment of
inertia tensor (see Appendix A).  The ``macroporosity'' $\poromacro$
is the measure of porosity put forward by
\citet[][Eq.\ 12]{Shen+Draine+Johnson_2008}.  The asymmetry parameter
$\Asymm$ is defined by \citet[][Eq.\ 3]{Draine_2024b}.  The
``stretch'' parameter $\Stretch$, a measure of whether the aggregate
shape is flattened or elongated, is defined by
\citet[][Eq.\ 3]{Draine_2024a}.

\begin{figure}
\begin{center}  
\includegraphics[angle=0,width=\fwidthc,
                 clip=true,trim=0.5cm 5.0cm 0.1cm 4.2cm]
{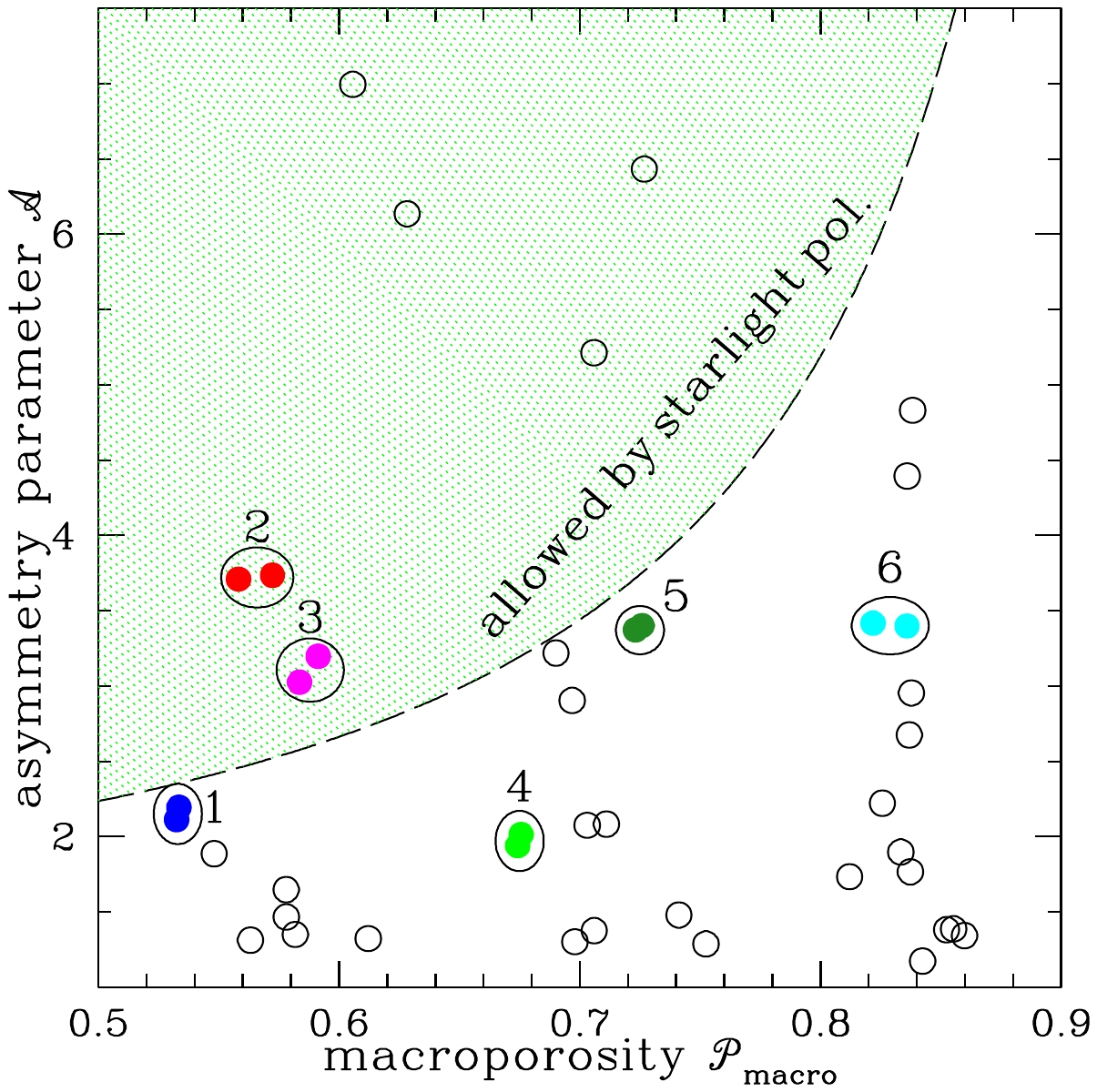}
\caption{\label{fig:sample}\footnotesize Irregular aggregates from
  \citet{Draine_2024a} on the $\poromacro-\Asymm$ plane.  Selected
  cases (see Table \ref{tab:targets}) are shown and identified.  The
  dashed curve is Eq.\ (19) from \citet{Draine_2024b}, the approximate
  boundary of the region in the $\poromacro-\Asymm$ plane allowed by
  the observed polarization of starlight.
  }
\end{center}
\end{figure}
\begin{table}
\begin{center}
\caption{\label{tab:targets} Targets}
{\footnotesize
\begin{tabular}{c c c c c r r r}
\hline
case & target\,$^a$      
    & $\poromacro\,^{b}$ & $\Asymm\,^{c}$ & $\Stretch\,^{d}$
    & $N_1$\,$^{e}$\, & $N_2$\,$^{e}$\, & $N_3$\,$^{e}$\, \\
\\
\hline
1 & BAM2trimA.256.2 & 0.533    & 2.195  & 1.814  & 153420 & 65103 & 37800\\
  & BAM2trimA.256.3 & 0.533    & 2.113  & 2.032  & 106205 & 61793 & 31709\\

  & prolate spheroid:\,1\,:\,0.348\,:\,0.348   & 0.53 & 2.151 &  2.151 
                                                  & 99992 & 51516 & 24965\\

\hline
2 & BAM2trimB.256.2 & 0.558      & 3.709 & 0.955  & 178617 & 125619 & 72935\\
  & BAM2trimB.256.3 & 0.573      & 3.735 & 1.362  & 55163  & 23504  & 12255\\

  & oblate spheroid:\,1\,:\,3.72\,:\,3.72 & 0.565  & 3.720 & 0.732 
                                                 & 104288 & 60532 & 31184\\

\hline

3 & BAM2trimB.256.1 & 0.584     & 3.025 & 0.807 & 231236 & 119352 & 61683\\
  & BAM2trimB.256.4 & 0.591     & 3.197 & 0.974 & 51260  & 29865  & 15372\\

  & oblate spheroid:\,1\,:\,3.11\,:\,3.11 & 0.59   & 3.11  & 0.743 
                                                 & 124864 & 72808 & 37316 \\
\hline
4 & BAM1trimA.256.1 & 0.676      & 2.014 & 1.698 & 231109 & 99109 & 57035 \\
  & BAM1trimA.256.3 & 0.674      & 1.939 & 1.706 & 58951  & 34288 & 17635 \\

  & prolate spheroid:\,1\,:\,0.382\,:\,0.382 & 0.675 & 1.982  & 1.982
                                                 & 91000 & 46932 & 22671 \\
\hline
5 & BAM1trimB.256.2 & 0.723      & 3.372 & 1.107 & 234224 & 147799 & 85718 \\
  & BAM1trimB.256.4 & 0.725      & 3.401 & 1.035 & 108484 & 62895  & 32376 \\

  & oblate spheroid:\,1\,:\,3.39\,:\,3.39  & 0.72 & 3.390 & 0.737
                                                  & 114608 & 66558 & 34272 \\
\hline
6 & BAtrimB.256.2 & 0.836      & 3.398 & 0.930  & 91070 & 43515 & 22375\\
  & BAtrimB.256.4 & 0.822      & 3.417 & 0.959  & 55960 & 32387 & 16668\\

  & oblate spheroid:\,1\,:\,3.41\,:\,3.41 & 0.830  & 3.410 & 0.737
                                                 & 113832 & 66182 & 34080  \\
\hline
\multicolumn{8}{l}{$a$~Target geometry from \cite{Draine_2024b}.}\\
\multicolumn{8}{l}{$b$~Macroporosity $\poromacro$ 
                   \citep[][Eq.\ 12]{Shen+Draine+Johnson_2008}.}\\
\multicolumn{8}{l}{$c$~Asymmetry parameter $\Asymm$ 
                   \citep[][Eq.\ 3]{Draine_2024b}.}\\
\multicolumn{8}{l}{$d$~Stretch parameter $\Stretch$
                   \citep[][Eq.\ 3]{Draine_2024a}.}\\
\multicolumn{8}{l}{$e$~Number of dipoles in DDA target realization.}\\
\end{tabular}
}
\end{center}
\end{table}


The values of $(\poromacro,\Asymm)$ for the 42 irregular aggregate
shapes studied by \citet{Draine_2024b} are shown in Figure
\ref{fig:sample}.  The shapes studied included a number of cases where
two different irregular aggregates happen to have similiar values of
$(\poromacro,\Asymm)$.  Six such cases are identified in Figure
\ref{fig:sample}, with parameters listed in Table \ref{tab:targets}.
The cases are numbered (1--6) in order of increasing $\poromacro$,
with $\poromacro$ ranging from 0.53 to 0.83.

We choose to study these pairs because differences in cross sections
between two aggregates with nearly the same $(\poromacro,\Asymm)$ will
reveal the degree to which factors other than $(\poromacro,\Asymm)$
may also be important.  Two of the cases (case 2 and case 3) fall in
the green zone of Figure \ref{fig:sample}, with values of
$(\poromacro,\Asymm)$ compatible with Eq.\ (19) from
\citet{Draine_2024b}, such that the aggregate (with suitable degree of
alignment) would be able to reproduce the observed starlight
polarization integral \citep{Draine+Hensley_2021c}.  However, case 1
and case 5 are both close to the estimated boundary (we will see below
that case 5 is allowed by starlight polarization).

\subsection{Spheroidal Analog}

Let $(a,b,b)$ be the semimajor axes of the spheroidal analog.  Its
volume is set to be
\beq
\frac{4\pi}{3} a b^2 = \frac{V_\solid}{1-\poromacro}
~~~.
\eeq
We require the spheroidal analog to have the same $\Asymm$ as the
irregular aggregate.  All shapes (other than a sphere) have
$\Asymm>1$.  For every $\Asymm>1$, there is both an oblate and a
prolate spheroid with the same $\Asymm$.  To approximate irregular
shapes, we use prolate spheroids when the stretch parameter
$\Stretch>1.5$, and oblate spheroids when $\Stretch<1.5$, with axial
ratios
\beq \label{eq:oblate}
b/a=
\begin{cases} \Asymm \hspace*{3.0cm}{\rm for~}\Stretch<1.5\\
              (2\Asymm^2-1)^{-1/2} \hspace*{1.0cm}{\rm for~}\Stretch>1.5
\end{cases}
\eeq
Thus the size and shape of the spheroidal analog are determined by
geometric properties of the aggregate: the solid volume $V_\solid$, the
macroporosity $\poromacro$, the asymmetry parameter $\Asymm$, and
$\sgn[\Stretch-1.5]$.  The dielectric function for the interior of the
spheroidal analog is calculated using an EMT with
$\epsilon_\solid=\epsilon_{\rm Ad}$ and $\fvac=\poromacro$.

The axial ratios used for the six cases are given in Table
\ref{tab:targets}.  Prolate spheroids are used for cases 1 and 4; for
the other 4 cases, oblate spheroids are used.

\subsection{Cross Sections for Extinction and Polarization}

The DDA\footnote{%
  We used the public-domain code DDSCAT 7.3.3, available at
  \url{http://ddscat.wikidot.com}.}
was used to calculate absorption and scattering for the irregular
aggregates and the spheroidal analogs with various orientations with
respect to the incoming polarized plane wave (see Appendix
\ref{app:DDA}).  For each target geometry (including the
spheroids),\footnote{%
  While more efficient computational methods exist for spheroids,
  e.g., the separation of variables method
  \citep{Voshchinnikov+Farafonov_1993}, accurate DDA calculations for
  spheroids take only a small fraction of the time required for the
  irregular targets that are the subject of this paper.}
DDA computations were carried out for three different numbers of
dipoles ($\Ndip=N_1$, $N_2$, $N_3$), with results extrapolated to
$\Ndip\rightarrow\infty$ and DDA uncertainties $\Delta Q^{\rm DDA}$
estimated using Equations (8,9) of \citet{Draine_2024a}.

We consider efficiency factors $Q_{\rm ext,ran}(\lambda)\equiv C_{\rm
  ext,ran}(\lambda)/\pi\aeff^2$ for extinction by randomly-oriented
aggregates and spheroids, and $Q_\polPSA(\lambda)\equiv
C_\polPSA(\lambda)/\pi\aeff^2$ for polarization by aggregates and
spheroids in ``perfect spinning alignment'' (PSA).\footnote{%
Spinning around $\bahat_1$ (the principal axis of largest moment of
inertia), with $\bahat_1$ perpendicular to the line-of-sight
\citep[see][]{Draine_2024a}.}

Table \ref{tab:results} lists selected quantities for each of the
aggregates, and for each of the spheroidal analogs (for the four
different EMTs).

To test the SAM, we consider aggregates with $\aeff=0.25\micron$.
This size is chosen because it is in the middle of the size range that
(1) dominates the extinction at visible wavelengths, (2) dominates the
polarization of starlight (and the polarized submm emission), and (3)
is characteristic of the grains that dominate the total dust mass.

\begin{figure}[t]
\begin{center}  
\includegraphics[angle=0,width=\fwidthb,
                 clip=true,trim=0.5cm 5.0cm 0.1cm 4.2cm]
{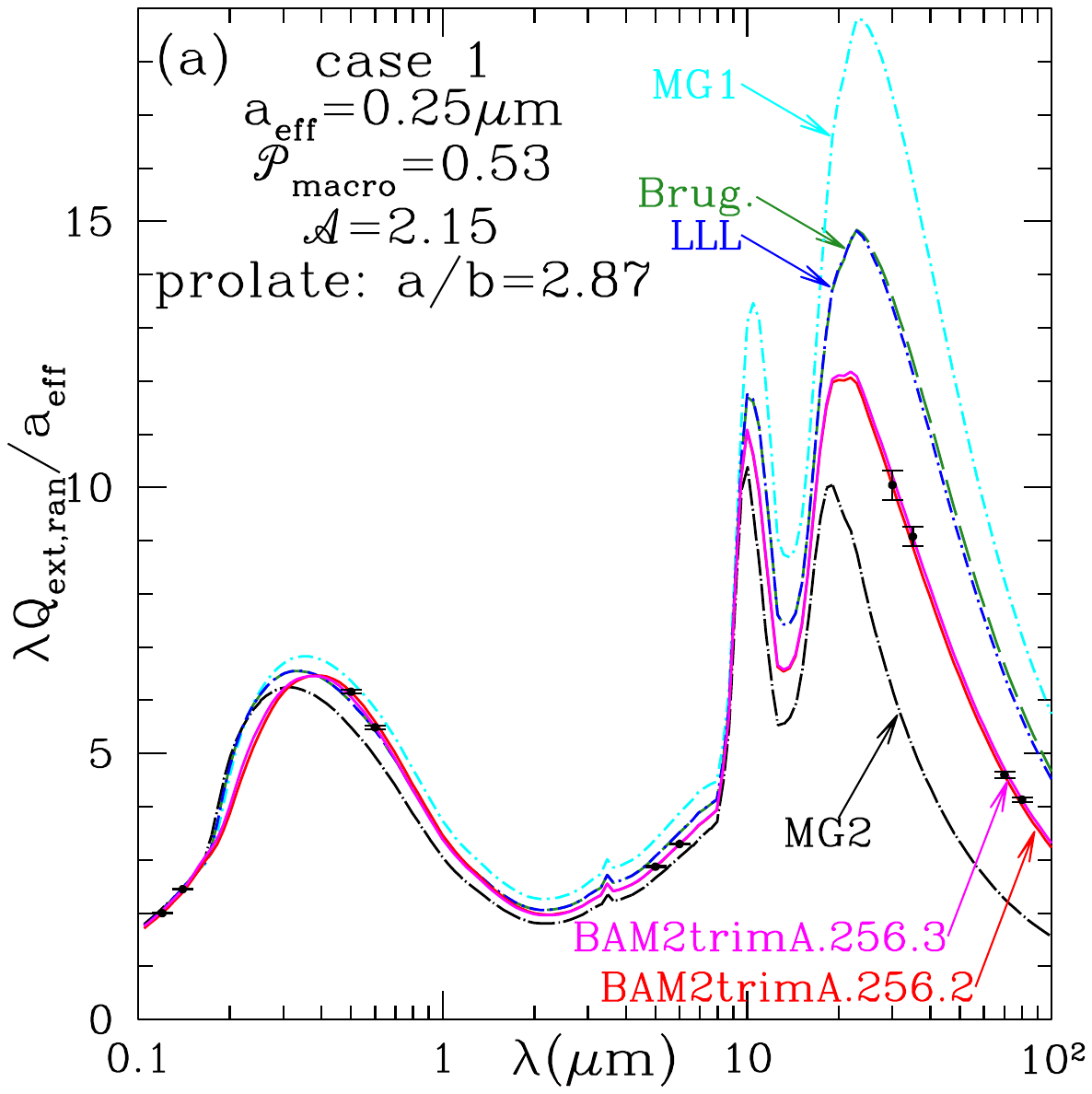}
\includegraphics[angle=0,width=\fwidthb,
                 clip=true,trim=0.5cm 5.0cm 0.1cm 4.2cm]
{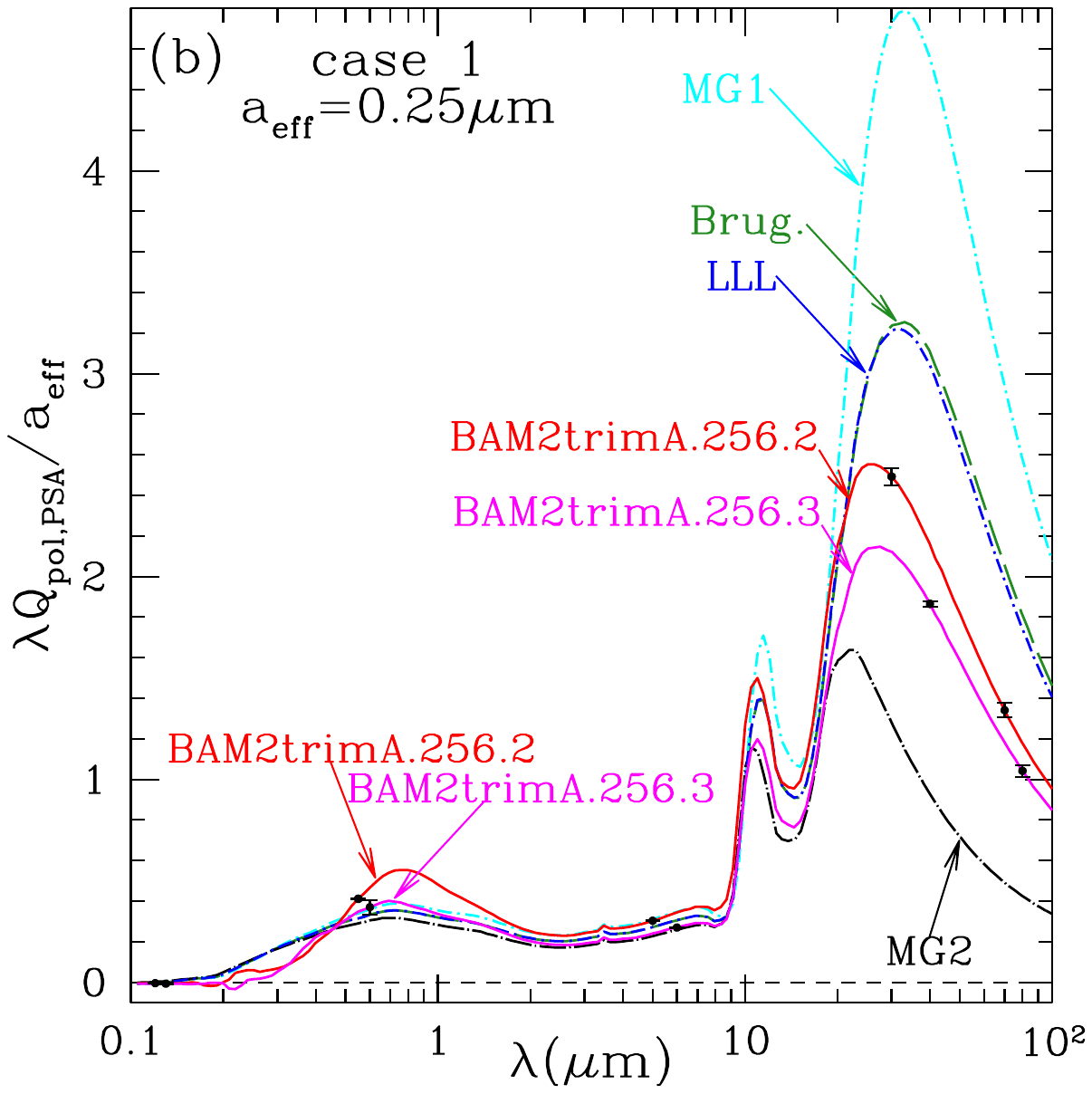}
\includegraphics[angle=0,width=\fwidthb,
                 clip=true,trim=0.5cm 5.0cm 0.1cm 4.2cm]
{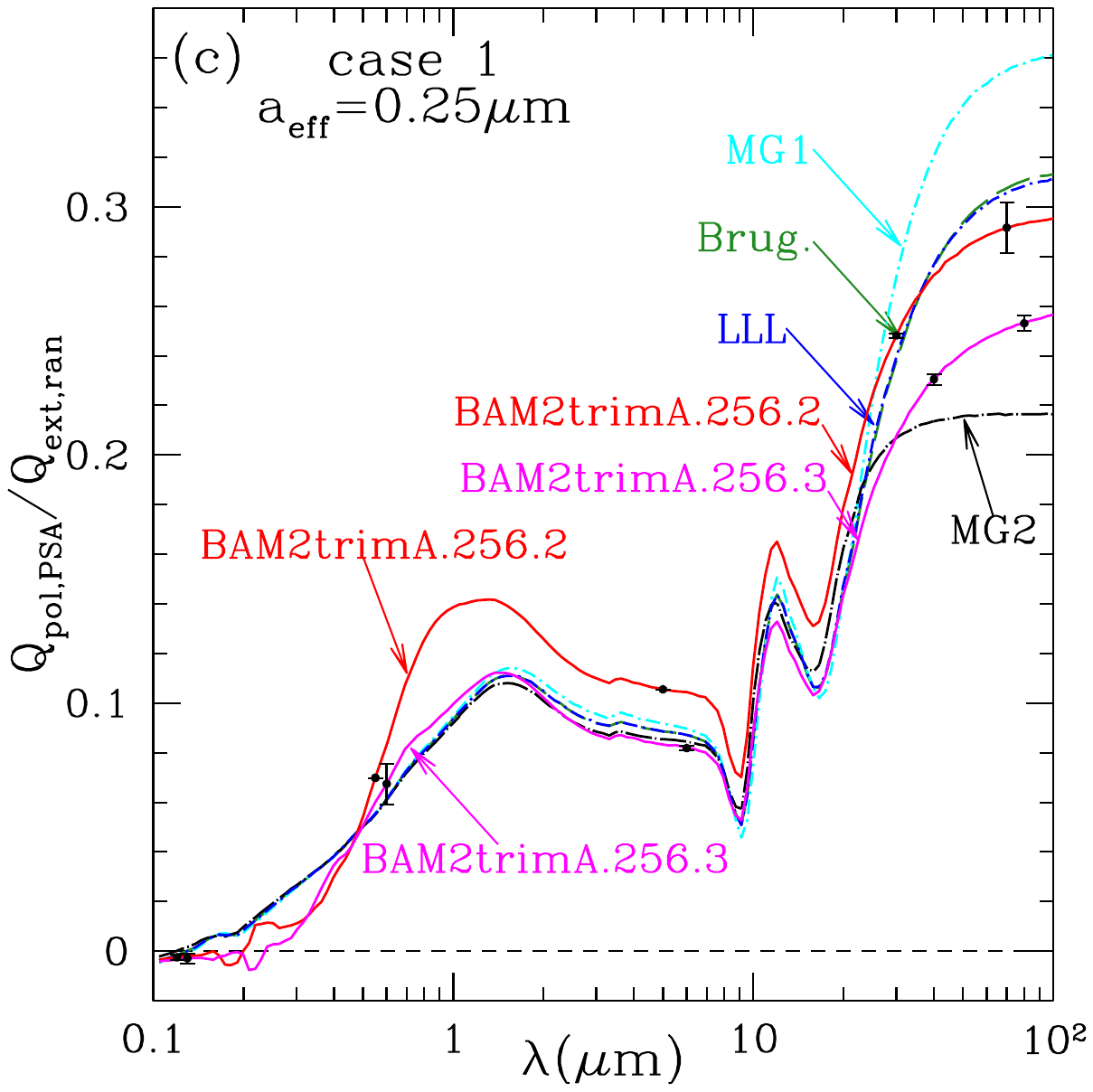}
\caption{\label{fig:Q_emt0.53}\footnotesize {\bf (a)} $\lambda
  Q_\extran/\aeff$, {\bf (b)} $\lambda Q_\polPSA/\aeff$, and
  {\bf (c)} $Q_\polPSA/Q_\extran$ for case 1
  irregular aggregates ($\poromacro\approx0.53$, $\Asymm\approx2.15$)
  with error bars at selected wavelengths indicating uncertainty in
  the DDA-calculated cross sections for the irregular aggregates.
  Also shown are SAM results using the Bruggeman (Brug.), Maxwell
  Garnett (MG1, MG2), and Landau-Lifshitz-Looyenga (LLL) EMTs.  The
  Bruggeman and LLL EMTs provide the best approximation for this case.
  }
\end{center}
\end{figure}
\begin{figure}
\begin{center}  
\includegraphics[angle=0,width=\fwidthb,
                 clip=true,trim=0.5cm 5.0cm 0.1cm 4.2cm]
{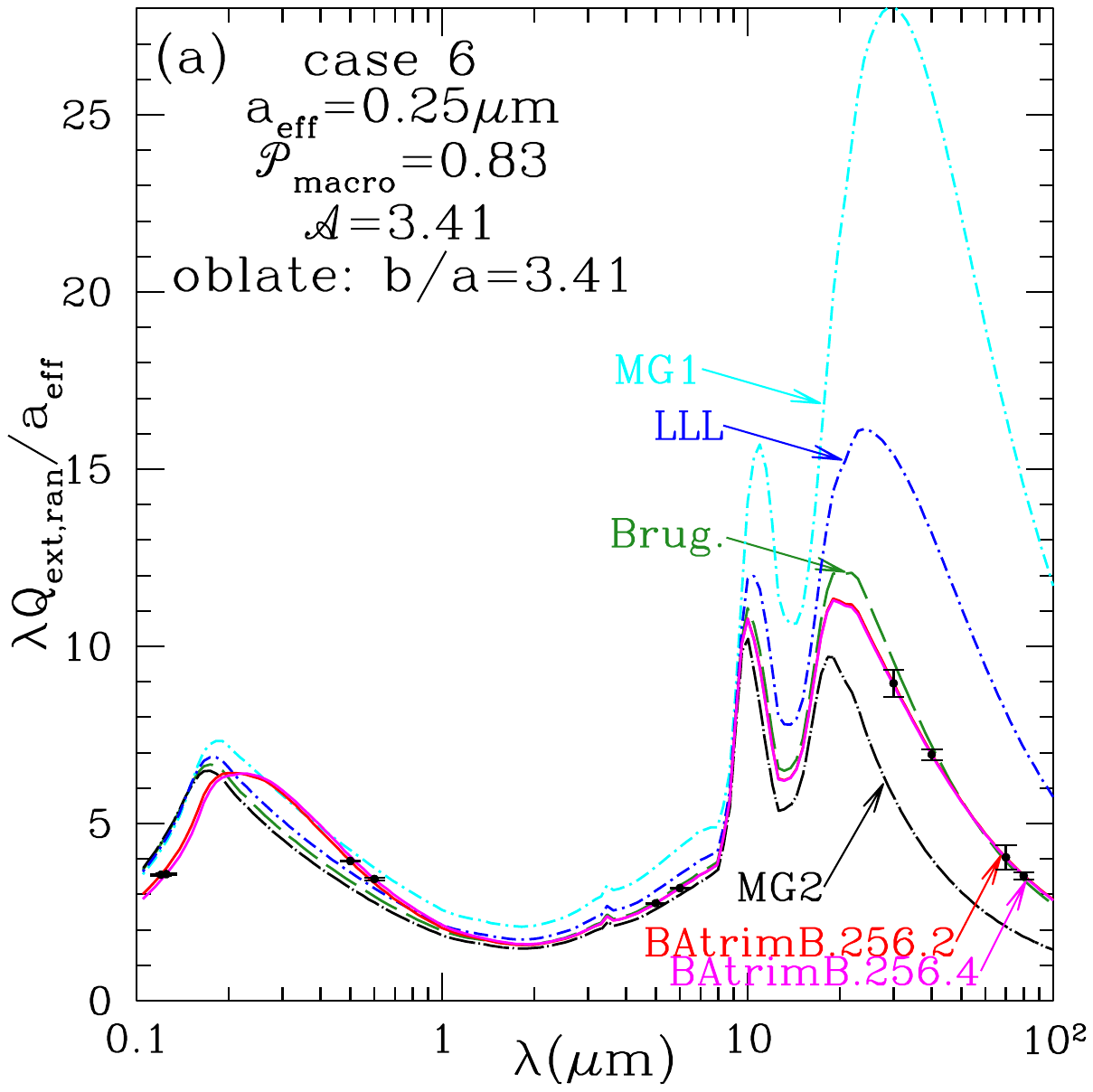}
\includegraphics[angle=0,width=\fwidthb,
                 clip=true,trim=0.5cm 5.0cm 0.1cm 4.2cm]
{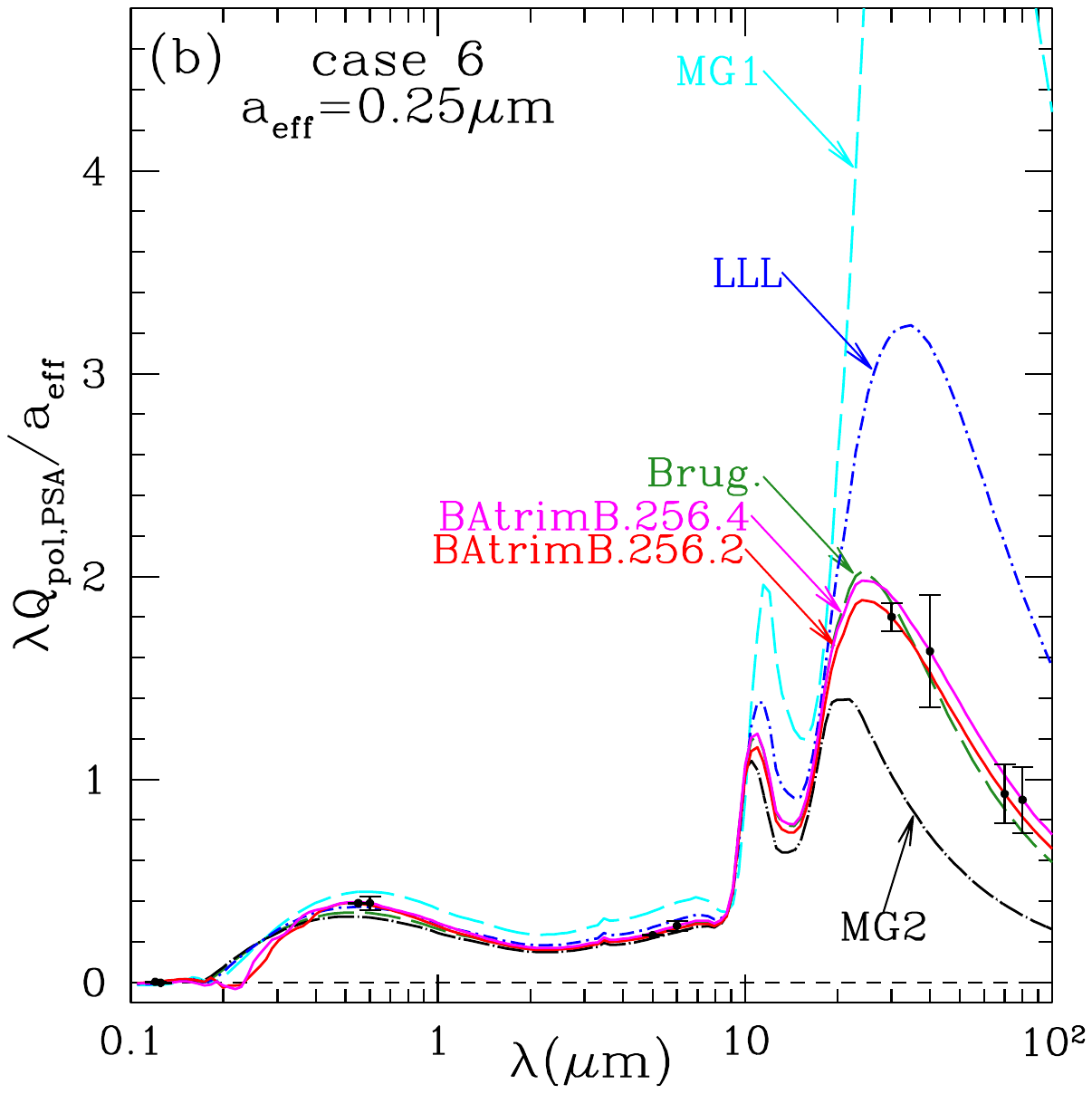}
\includegraphics[angle=0,width=\fwidthb,
                 clip=true,trim=0.5cm 5.0cm 0.1cm 4.2cm]
{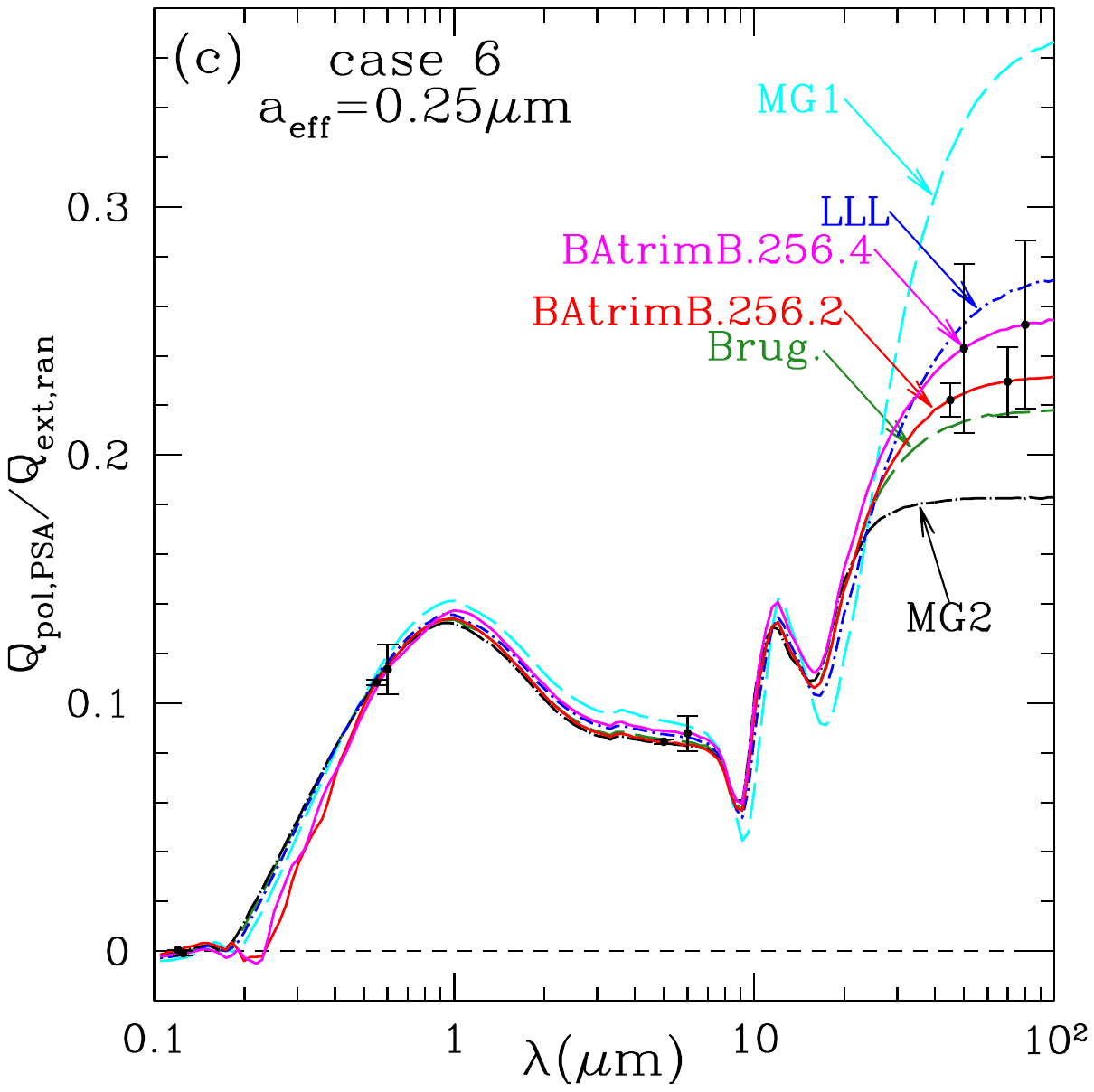}
\caption{\label{fig:Q_emt0.83}\footnotesize Same as Figure
  \ref{fig:Q_emt0.53} but for case 6 aggregates
  ($\poromacro\approx0.83$, $\Asymm\approx3.41$).  The Bruggeman EMT
  comes closest to the true results.
  }
\end{center}
\end{figure}

\subsubsection{Case 1 Aggregates: Moderate Porosity}

Figure \ref{fig:Q_emt0.53} shows $Q_\extran$ and $Q_\polPSA$
calculated for the two ``case 1'' aggregates
$(\poromacro\approx0.53,\Asymm\approx2.15)$ for size
$\aeff=0.25\micron$ and wavelengths $\lambda$ from $0.1\micron$ to
$100\micron$.  For random orientations, the BAM2trimA.256.2 and
BAM2trimA.256.3 aggregates have nearly identical extinction cross
sections (Figure \ref{fig:Q_emt0.53}a), although the polarization
properties for perfect spinning alignment differ noticeably for
$\lambda \gtsim 0.6\micron$ (Figure \ref{fig:Q_emt0.53}b).  The DDA
uncertainties in the calculated $Q_\extran$ and $Q_\polPSA$ are shown at
selected wavelengths.

Figure \ref{fig:Q_emt0.53} also shows SAM results for case 1 (2.87:1
prolate spheroids).  All four EMTs provide a good approximation for
$Q_\extran(\lambda)$ for $\lambda\ltsim 10\micron$, but for $\lambda >
10\micron$ the different EMTs predict quite different values of
$Q_\extran$, reflecting the variation among the different EMTs for
$\lambda > 10\micron$ (see Figure \ref{fig:eps_emt}a).  For this value
of $\poromacro$, the Bruggeman and LLL EMTs give very similar results
(as expected from the similarity of the Bruggeman and LLL dielectric
functions in Figure \ref{fig:eps_emt}a).  For $\lambda < 10\micron$,
$Q_\extran(\lambda)$ calculated for spheroids with the Bruggeman (or
LLL) EMT is within $\sim$20\% of the exact result for the case 1
irregular aggregates, with the largest fractional errors arising at UV
wavelengths $\lambda < 0.3\micron$.  At long wavelengths $\lambda
\gtsim 30\micron$ the SAM with either Bruggeman or LLL EMT
overestimates $Q_\extran$ by $\sim$40\%.  Over the full range of
wavelengths, the Bruggeman and LLL EMTs are usually closest to the
``exact'' results for $Q_\extran$.

The polarization cross sections calculated with the SAM (see Figure
\ref{fig:Q_emt0.53}b) are less accurate. For $\lambda\gtsim 50\micron$
the error in the Bruggeman or LLL EMT estimate for $Q_\polPSA$ can be
as large as $\sim$40\% (see Figure \ref{fig:Q_emt0.53}a).  Note,
however, that despite having very similar values of $\Asymm$, the two
case 1 aggregates also differ significantly from one another --
$Q_\polPSA$ is sensitive to structural details.

The ratio of polarization to extinction, $Q_\polPSA/Q_\extran$, is
plotted in Figure \ref{fig:Q_emt0.53}c.  The SAM using the Bruggeman
EMT reproduces the average of the two irregular aggregates to within
$\sim$10\% for $\lambda > 0.4\micron$.

\subsubsection{Case 6 Aggregates: High Porosity}

Figure \ref{fig:Q_emt0.83} shows cross sections calculated for the
highest-porosity case: case 6, with $(\poromacro,\Asymm)=(0.83,3.41)$.
The two case 6 aggregates (BAtrimB.256.2 and BAtrimB.256.3) have very
similar cross sections for total and polarized extinction over the
full wavelength range.  The oblate spheroids intended to emulate them
provide a fairly good approximation at optical wavelengths, but for
$\lambda > 10\micron$ the extinction and polarization cross sections
calculated with different EMTs differ greatly (see Figure
\ref{fig:Q_emt0.83}a), with very large differences in the
far-infrared, reflecting the large difference between the different
dielectric functions (see Figure \ref{fig:eps_emt}b).  The Bruggeman
EMT provides a good approximation for both $Q_\extran$ and
$Q_\polPSA$.  Figure \ref{fig:Q_emt0.83}c shows very good agreement
between $Q_\polPSA/Q_\extran$ calculated directly for the aggregates,
and calculated with the SAM using the Bruggeman EMT.

\begin{table}
\begin{center}
\caption{\label{tab:results} Selected Results}
{\footnotesize
\begin{tabular}{c c c c c c c c}
\hline
case & target$^a$      
    &  method 
    & $\aeffp\,^{a}$       
    & $\Phi_\PSA\,^{b}$ 
    & $\sigmap\,^{c}$
    & $C_\extran(100\micron)/V_\solid\,^{d}$
    & $\pfirmax\,^{e}$
\\
    &                    
    &                    
    & ($\mu$m)
    &
    &
    & $(\cm^{-1})$
    &
\\
\hline
1 & BAM2trimA.256.2    
    & DDA    
    & $0.214$ 
    & $0.579\pm0.001$
    & $0.548\pm0.016$   
    & $242.0\pm1.8$
    & $0.222\pm0.010$ \\
  & BAM2trimA.256.3 
    & DDA    
    & $0.207$ 
    & $0.453\pm0.011$
    & $0.502\pm0.072$
    & $248.4\pm2.8$
    & $0.191\pm0.004$ \\

  &sph.\,1\,:\,0.348\,:\,0.348 
  & Brug.\,EMT 
  & $0.280$ 
  & $0.467\pm0.005$
  & $0.612\pm0.003$  
  & $349.6\pm1.0$
  & $0.237\pm0.003$ \\

  & ''
  & LLL EMT
  & $0.280$
  & $0.468\pm0.005$
  & $0.612\pm0.003$
  & $338.6\pm2.2$
  & $0.235\pm0.003$ \\

  & ''
  & MG1 EMT
  & $0.269$ 
  & $0.513\pm0.005$
  & $0.611\pm0.002$
  & $430.8\pm2.8$
  & $0.276\pm0.004$ \\

  & ''
  & MG2 EMT
  & $0.295$ 
  & $0.415\pm0.004$
  & $0.613\pm0.003$
  & $116.0\pm0.3$
  & $0.160\pm0.001$\\
\hline
2 & BAM2trimB.256.2
  & DDA
  & $0.210$
  & $1.220\pm0.018$
  & $0.467\pm0.008$
  & $267.2\pm8.3$
  & $0.209\pm0.004$\\

  & BAM2trimB.256.3
  & DDA
  & $0.207$
  & $1.065\pm0.024$
  & $0.471\pm0.003$
  & $274\pm18$
  & $0.224\pm0.004$\\

  & sph.\,1\,:\,3.72\,:\,3.72
  & Brug.\,EMT
  & $0.240$
  & $1.265\pm0.028$
  & $0.526\pm0.001$
  & $386.0\pm5.1$
  & $0.210\pm0.006$\\

  & ''
  & LLL EMT
  & $0.239$
  & $1.398\pm0.031$
  & $0.517\pm0.001$
  & $386.4\pm4.8$
  & $0.209\pm0.006$\\
 
  & ''
  & MG1 EMT
  & $0.230$
  & $1.398\pm0.031$
  & $0.526\pm0.001$
  & $516.1\pm7.4$
  & $0.207\pm0.006$\\

  & ''
  & MG2 EMT
  & $0.251$
  & $1.135\pm0.025$
  & $0.527\pm0.002$
  & $122.1\pm0.8$
  & $0.187\pm0.004$\\
\hline

3 & BAM2trimB.256.1
  & DDA
  & $0.204$
  & $0.990\pm0.005$
  & $0.471\pm0.001$
  & $260\pm13$
  & $0.225\pm0.003$\\

  & BAM2trimB.256.4
  & DDA
  & $0.205$
  & $0.960\pm0.018$
  & $0.474\pm0.012$
  & $269\pm20$
  & $0.230\pm0.001$\\

  & sph.\,1\,:\,3.11\,:\,3.11
  & Brug.\,EMT
  & $0.234$
  & $1.082\pm0.027$
  & $0.522\pm0.002$
  & $362.7\pm2.9$
  & $0.222\pm0.006$\\ 

  & ''
  & LLL EMT
  & $0.233$
  & $1.095\pm0.028$
  & $0.522\pm0.002$
  & $373.4\pm2.0$
  & $0.222\pm0.006$\\

  & ''
  & MG1 EMT
  & $0.223$
  & $1.210\pm0.030$
  & $0.521\pm0.000$
  & $504.7\pm5.4$
  & $0.221\pm0.009$\\

  & ''
  & MG2 EMT
  & $0.246$
  & $0.971\pm0.023$
  & $0.523\pm0.003$
  & $117.0\pm0.6$
  & $0.194\pm0.004$\\
\hline
4 & BAM1trimA.256.1
  & DDA
  & $0.242$
  & $0.358\pm0.047$
  & $0.597\pm0.065$
  & $242.1\pm5.5$
  & $0.146\pm0.017$\\

  & BAM1trimA.256.3
  & DDA
  & $0.268$
  & $0.291\pm0.085$
  & $0.635\pm0.025$
  & $250\pm12$
  & $0.136\pm0.010$\\

  & sph.\,1\,:\,0.382\,:\,0.382 
  & Brug.\,EMT
  & $0.323$
  & $0.284\pm0.001$
  & $0.638\pm0.007$
  & $307.1\pm0.3$
  & $0.140\pm0.001$\\

  & ''
  & LLL EMT
  & $0.319$
  & $0.293\pm0.001$
  & $0.637\pm0.007$
  & $374.8\pm0.3$
  & $0.153\pm0.001$\\

  & ''
  & MG1 EMT
  & $0.301$
  & $0.335\pm0.001$
  & $0.635\pm0.006$
  & $558.1\pm1.2$
  & $0.200\pm0.001$\\

  & ''
  & MG2 EMT
  & $0.338$
  & $0.255\pm0.001$
  & $0.640\pm0.007$
  & $108.3\pm0.2$
  & $0.096\pm0.001$\\
\hline
5 & BAM1trimB.256.2
  & DDA       
  & $0.240$
  & $0.748\pm0.005$
  & $0.500\pm0.009$
  & $251.0\pm7.0$
  & $0.255\pm0.010$\\

  & BAM1trimB.256.4
  & DDA
  & $0.241$
  & $0.744\pm0.058$
  & $0.508\pm0.020$
  & $261.0\pm9.2$
  & $0.261\pm0.001$\\

  & sph.\,1\,:\,3.39\,:\,3.39
  & Brug.\,EMT
  & $0.271$
  & $0.766\pm0.010$
  & $0.538\pm0.001$   
  & $295.0\pm0.4$
  & $0.243\pm0.003$\\

  & ''
  & LLL EMT
  & $0.303$
  & $0.340\pm0.001$
  & $0.637\pm0.007$   
  & $435.0\pm0.4$
  & $0.153\pm0.001$\\

  & ''  
  & MG1 EMT
  & $0.251$
  & $0.916\pm0.012$
  & $0.537\pm0.001$
  & $677.1\pm2.9$
  & $0.269\pm0.003$\\

  & ''  
  & MG2 EMT
  & $0.282$
  & $0.698\pm0.009$
  & $0.540\pm0.002$
  & $110.9\pm0.1$
  & $0.208\pm0.003$\\
\hline
6 & BAtrimB.256.2
  & DDA
  & $0.274$
  & $0.447\pm0.002$
  & $0.527\pm0.004$
  & $214\pm21$
  & $0.171\pm0.013$\\

  & BAtrimB.256.4
  & DDA
  & $0.270$
  & $0.434\pm0.012$
  & $0.520\pm0.007$
  & $206\pm38$
  & $0.163\pm0.012$\\

  & sph.\,1\,:\,3.41\,:\,3.41  
  & Brug.\,EMT
  & $0.326$
  & $0.452\pm0.007$
  & $0.566\pm0.001$
  & $202.9\pm0.1$
  & $0.161\pm0.002$\\

  & ''
  & LLL EMT
  & $0.316$
  & $0.488\pm0.007$
  & $0.567\pm0.006$   
  & $431.2\pm0.6$
  & $0.202\pm0.003$\\

  & ''
  & MG1 EMT
  & $0.294$
  & $0.583\pm0.008$
  & $0.573\pm0.014$
  & $872.3\pm8.3$
  & $0.281\pm0.003$\\

  & '' & MG2 EMT & $0.336$ & $0.422\pm0.006$ & $0.568\pm0.006$ &
$106.5\pm0.5$ & $0.134\pm0.002$\\
 \hline \multicolumn{8}{l}{$a$~Size
  $\aeff$ such that $\lambdap=0.567\micron$
  \citep[see][]{Draine_2024a}.}\\
 \multicolumn{8}{l}{$b$~Starlight
  polarization efficiency integral (Equation \ref{fig:PhiPSA}).}\\
 \multicolumn{8}{l}{$c$~Starlight
  polarization width parameter
  \citep[see][]{Draine+Hensley_2021c}.}\\
 \multicolumn{8}{l}{$d$~$\Cext/V_\solid\approx\Cabs/V_\solid$
  at $\lambda=100\micron$ for randomly oriented
  grains.}\\
 \multicolumn{8}{l}{$e$~FIR polarization fraction for
  perfect spinning alignment \citep[see][Eq.\, 8]{Draine_2024a}.}\\

\end{tabular}
}
\end{center}
\end{table}

\subsection{Accuracy of the Spheroidal Analog Method with Bruggeman EMT}

\begin{figure}
\begin{center}  
\includegraphics[angle=0,width=\fwidthc,
                 clip=true,trim=0.5cm 5.0cm 0.1cm 4.2cm]
{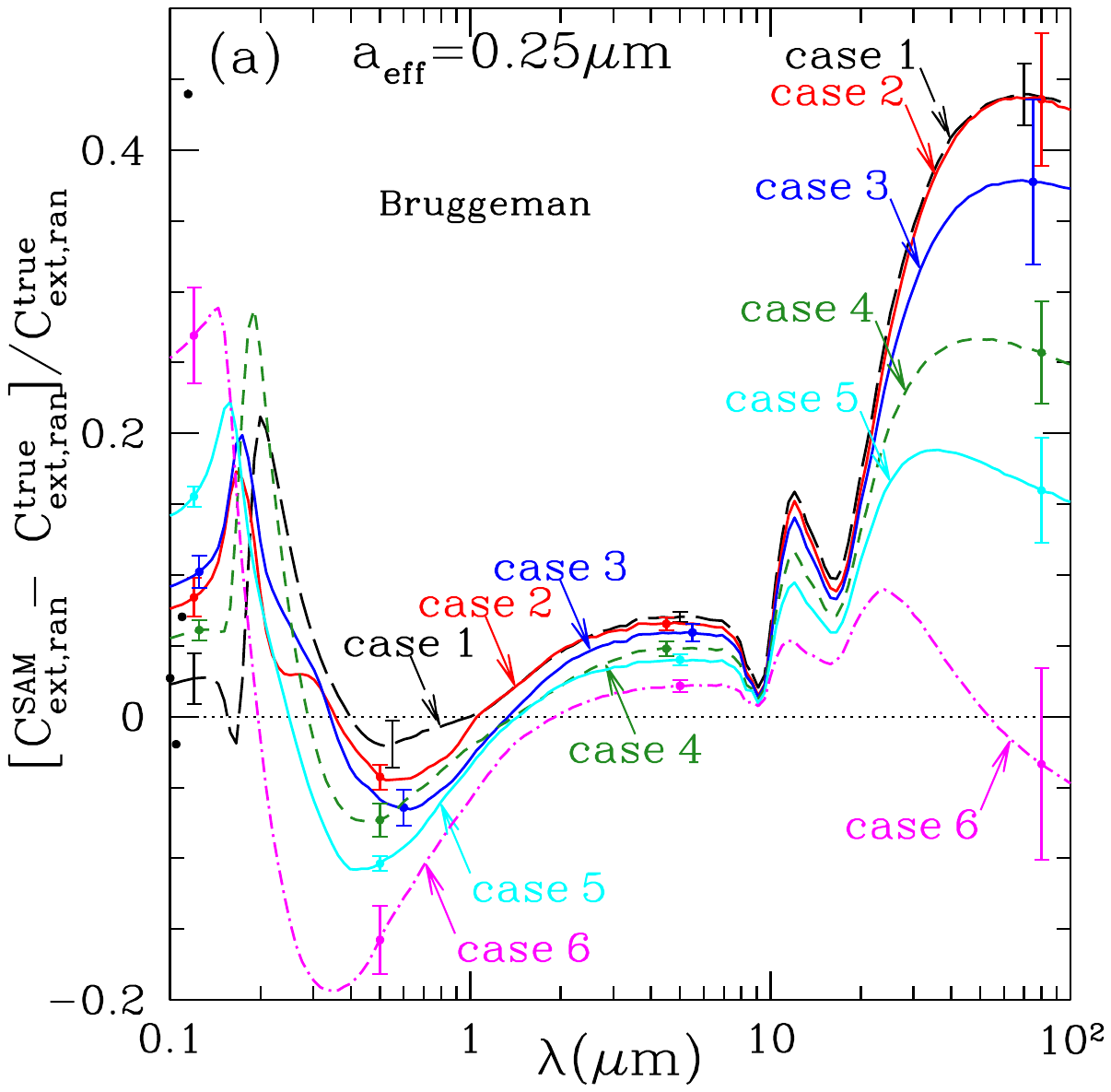}
\includegraphics[angle=0,width=\fwidthc,
                 clip=true,trim=0.5cm 5.0cm 0.1cm 4.2cm]
{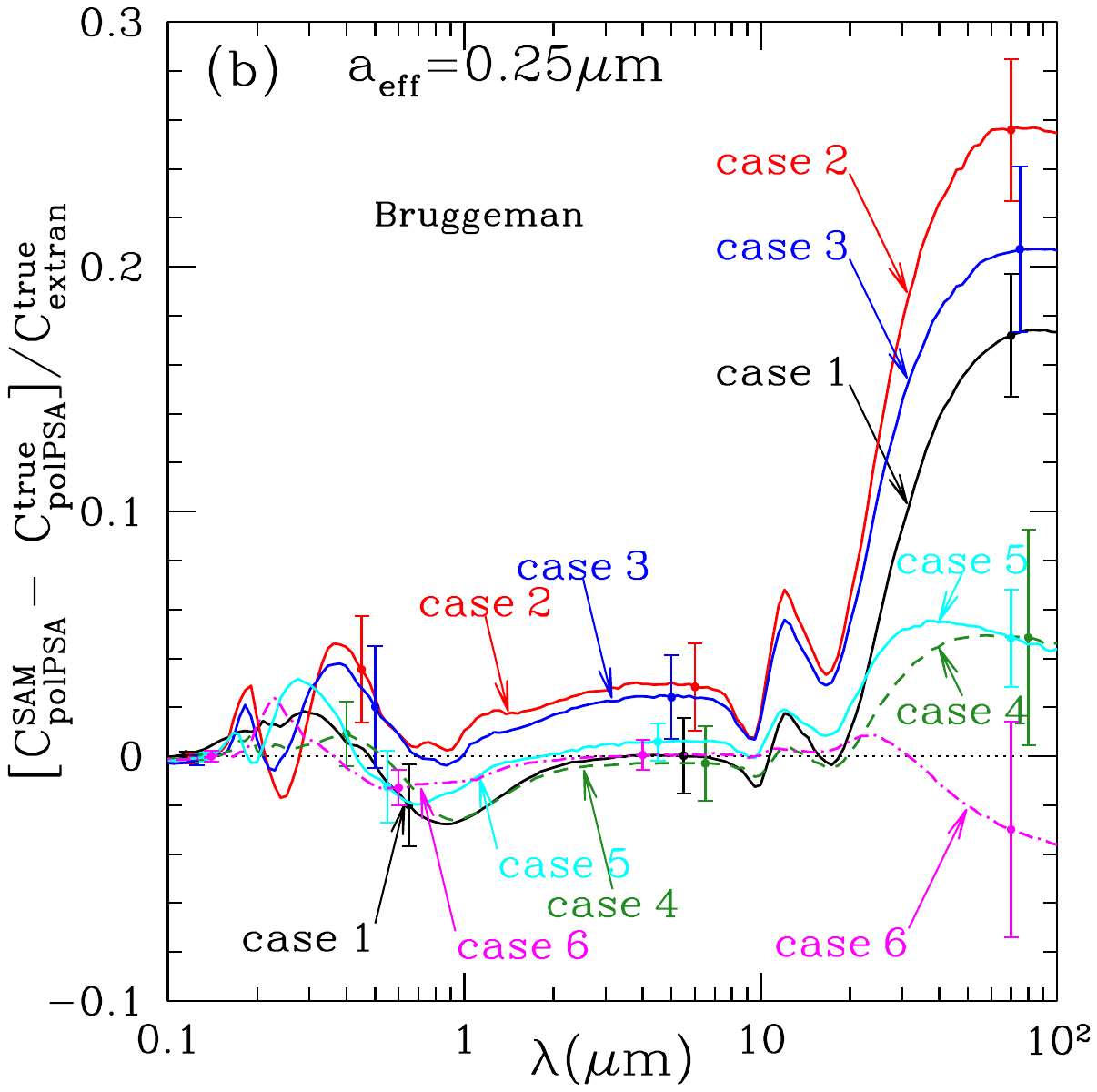}
\caption{\label{fig:err}\footnotesize
  Test of approximating irregular grains by spheroidal analogs using the
  Bruggemen EMT.  {\bf (a)} Fractional error in extinction cross sections
  for randomly-oriented $\aeff=0.25\micron$ irregular grains (cases
  1--6). For $0.2\micron \ltsim \lambda \ltsim 20\micron$, the
  fractional errors are less than 20\%.  {\bf (b)} Error in polarization
  cross section for $\aeff=0.25\micron$ irregular grains in perfect
  spinning alignment, relative to total extinction cross section for
  random alignment.  For each case, uncertainties arising from use of
  the DDA, and from real variation among irregular aggregates within
  the same ``case'' (see Equation \ref{eq:u(phi)}) are shown as
  ``error bars'' at selected wavelengths.
  }
\end{center}
\end{figure}

From this point forward, discussion of the accuracy of the SAM will
assume use of the Bruggeman EMT, as this gives the best overall results.

Consider irregular aggregates in some ``case'' characterized by
($\poromacro$, $\Asymm$, $\sgn[\Stretch-1.5]$).  Let
$\phi_j^\DDA(\aeff,\lambda)$ be some optical property (e.g.,
$Q_\extran$) computed directly with the DDA for one example $j$ of the
case.  If $K$ different examples of the case have been studied ($K=2$
in the present work), the best estimate of the ``true'' value
$\phi^{\rm true}$ is
\beq
\phi^{\rm true} ~\approx~
\langle\phi\rangle ~\equiv~
\frac{1}{K} \sum_{j=1}^K \phi_j^{\DDA}
~~~.
\eeq
where each of the $\phi_j^{\DDA}$ is obtained by extrapolating
$\Ndip\rightarrow\infty$.

The extrapolation $\Ndip\rightarrow\infty$ has an estimated
uncertainty $\Delta\phi_j^\DDA$ obtained from Equation (9) of
\citet{Draine_2024a}.  In addition, the $\phi_j^\DDA$ differ from
example to example.  The overall estimated uncertainty in the ``true''
value is
\beq \label{eq:u(phi)}
\Delta\phi = 
\left[
  \frac{1}{K(K-1)}\sum_{j=1}^K 
\left(\phi_j^\DDA-\langle\phi\rangle\right)^2 +
  \frac{1}{K}\sum_{j=1}^K (\Delta \phi_j^\DDA)^2 
\right]^{1/2}
~~~.
\eeq
Uncertainties $\pm \Delta Q_\extran/Q_\extran$ and $\pm \Delta
Q_\polPSA/Q_\extran$ are shown as ``error bars'' at selected
wavelengths in Figure \ref{fig:err}.  Note that they differ from the
``error bars'' in Figure \ref{fig:Q_emt0.53} and \ref{fig:Q_emt0.83}
in that they also include the variation between the two different
aggregates in each case.

Figure \ref{fig:err}a shows, for each of the 6 cases, the fractional
error in the SAM estimate for $Q_\extran$ when the Bruggeman EMT is
used.  For all 6 cases the fractional errors approach or exceed
$\sim$20\% at FUV wavlengths $\lambda \ltsim 0.2\micron$ -- this is
perhaps not surprising, given that the $\aeff=0.25\micron$ aggregates
under consideration are composed of 256 spheres with radii
$0.039\micron$ -- these individual spheres have sizes that are not
small compared to $\lambda/2\pi$, rendering the use of effective
medium theory questionable.  Nevertheless, the SAM remains a useful
approximation, even with errors reaching $\sim$20\%.

At wavelengths $\sim$$0.5\micron$, the errors shown in Figure
\ref{fig:err}a increase with increasing porosity, from case 1 to case
6.  However, even for case 6 the errors do not exceed 20\% for
$0.25\micron<\lambda < 20\micron$.

At wavelengths $\lambda > 20\micron$, the SAM errors become larger for
cases 1--4, with $\poromacro\in[0.53,0.68]$.  Interestingly, the SAM
does well for case 6, with the highest porosity $\poromacro=0.83$,
with errors $<10\%$ for $\lambda > 1\micron$.

Because $Q_\polPSA$ passes through zero, instead of discussing the
fractional error, we instead plot the error in $Q_\polPSA$ relative to
$Q_\extran$ in Figure \ref{fig:err}b.  By this measure, the errors in
$Q_\polPSA$ are modest for $\lambda < 20\micron$, but for cases 1--3
($\poromacro\in[0.53,0.59]$) the errors brecome substantial for
$\lambda \gtsim 30\micron$.  However, for the highest porosity cases
4--6 ($\poromacro\in[0.68,0.83]$), the SAM performs well for
polarization at all wavelengths.

\subsection{Starlight Polarization Efficiency Integral}

\begin{figure}
\begin{center}  
\includegraphics[angle=0,width=\fwidth,
                 clip=true,trim=0.5cm 5.0cm 0.1cm 4.2cm]
{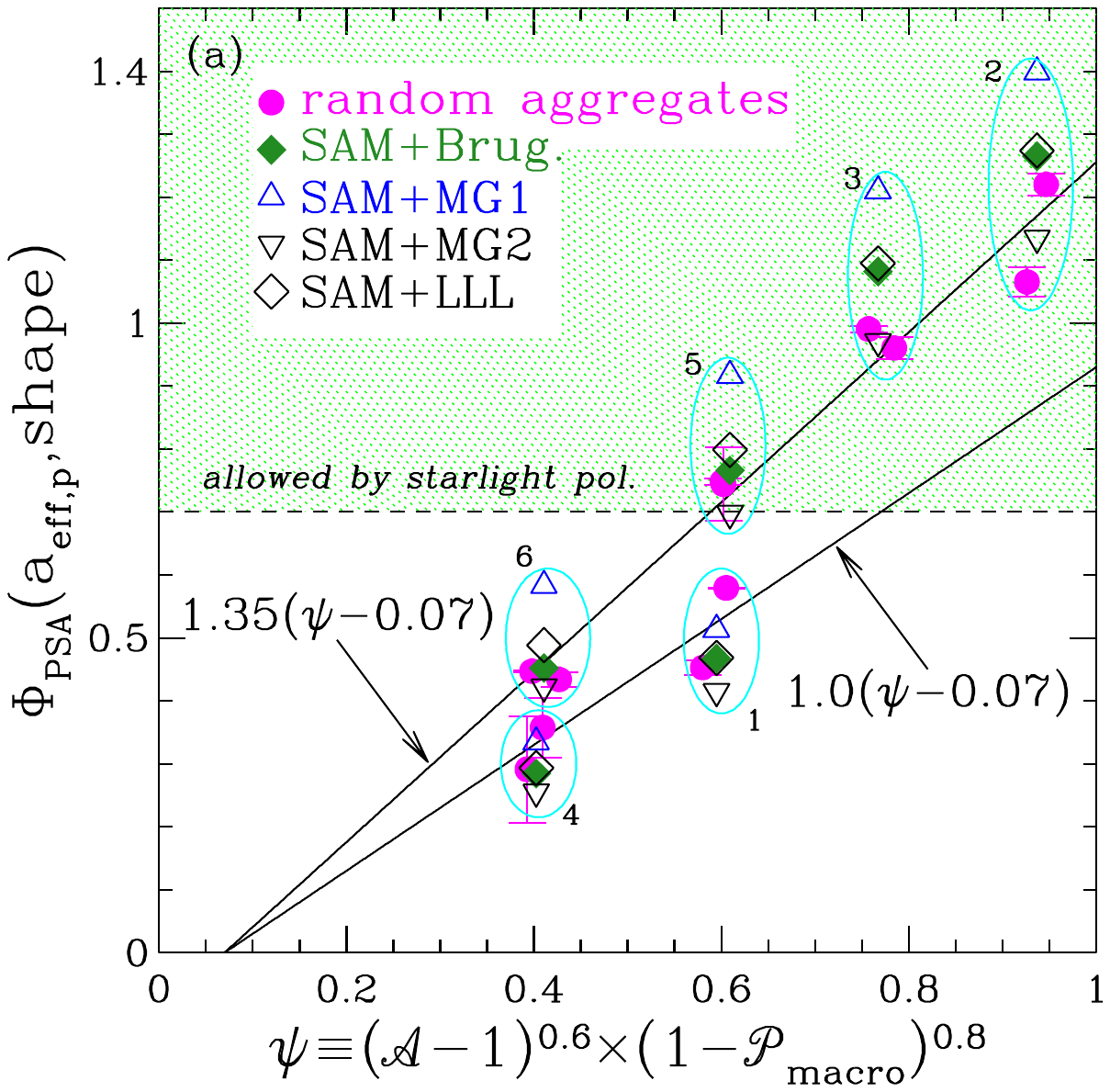}
\includegraphics[angle=0,width=\fwidth,
                 clip=true,trim=0.5cm 5.0cm 0.1cm 4.2cm]
{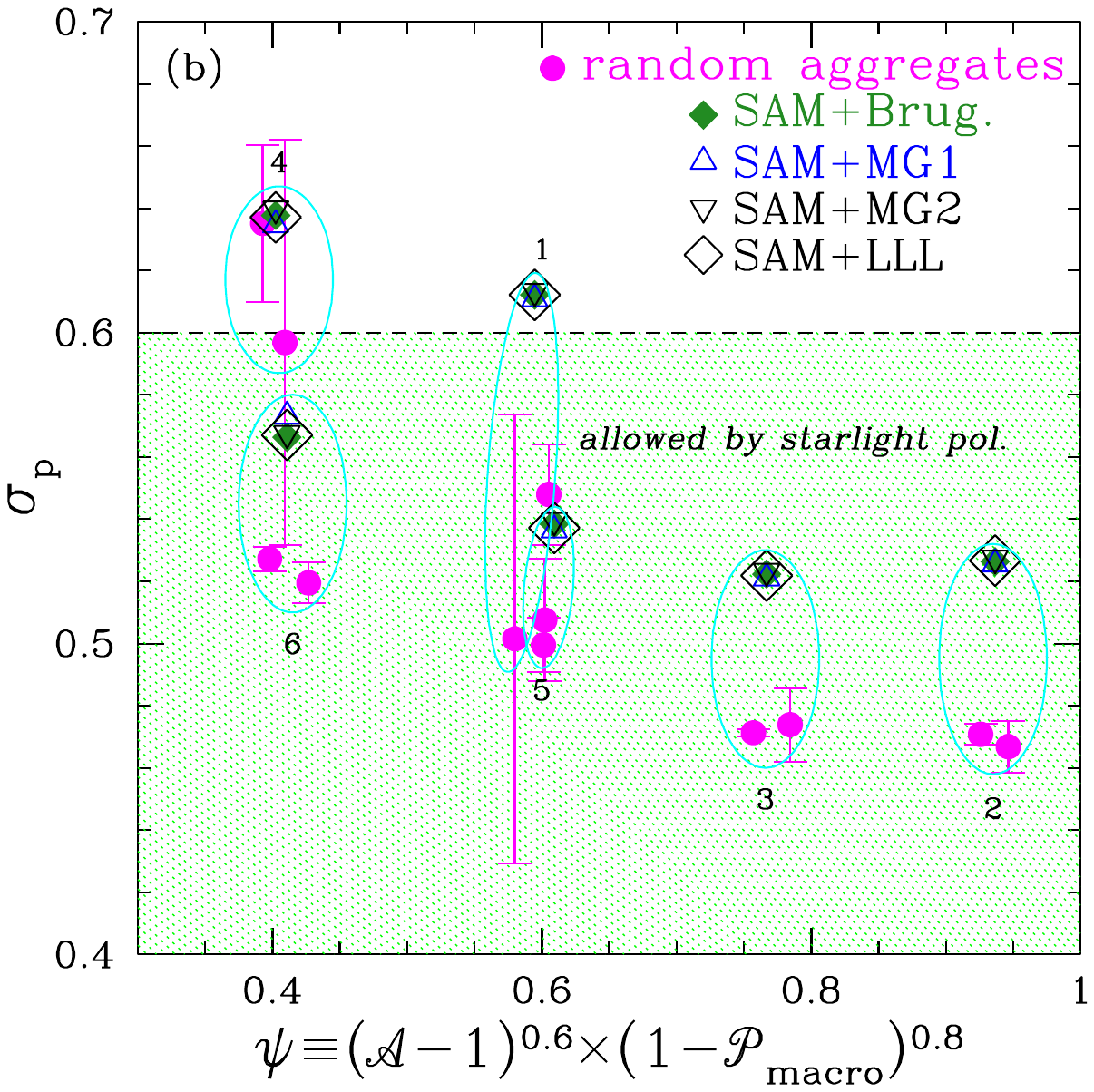}
\caption{\label{fig:PhiPSA}\footnotesize
  {\bf(a)} Starlight polarization efficiency integral $\Phi_\PSA$ and
  {\bf(b)} Starlight polarization width parameter $\sigmap$, evaluated
  using the DDA for twelve different random aggregates (filled
  circles) versus the parameter $\psi$, for the six cases in Table
  \ref{tab:targets}.  For each aggregate, error bars indicate the
  uncertainties of the DDA calculation.  Cases are indicated by the
  cyan ellipses (labeled by case number).  Also shown are $\Phi_\PSA$
  and $\sigmap$ calculated for spheroidal analogs, using four different
  EMTs.  Solid lines in {\bf (a)} are Eq.\ (\ref{eq:oblate trend}) and
  (\ref{eq:prolate trend}) for flattened (``oblate'') and elongated
  (``prolate'') shapes, respectively.  The dashed line in {\bf (b)}
  shows the upper limit to $\sigma_p$ allowed by observations
  \citep[see][]{Draine_2024a}.
  }
\end{center}
\end{figure}

The ``starlight polarization efficiency integral''
\citep{Draine+Hensley_2021c}
\beq \label{eq:PhiPSA}
\Phi_\PSA(\aeff) \equiv \frac{1}{V_\solid}
\int_{0.15\mu{\rm m}}^{2.5\mu{\rm m}}C_\polPSA(\aeff,\lambda) d\lambda
\eeq
is a useful dimensionless quantity to determine whether a candidate
grain shape is compatible with the observed polarization of starlight
in the diffuse ISM.  \citet{Draine+Hensley_2021c} showed that the
grains dominating interstellar extinction in the visible must have
$\Phi_\PSA(\aeffp)\gtsim0.7$ to be consistent with the observed
polarization of starlight.

We evaluate $\Phi_\PSA(\aeffp)$ for the different aggregates and their
spheroidal analogs, where, for each shape, $\aeffp$ is the size for
which the effective wavelength of polarization $\lambdap$
\citep[see][Eq.\ 39]{Draine+Hensley_2021c} matches the corresponding
wavelength $\lambdap=0.567\micron$ for the typical observed starlight
polarization profile -- this size is representative of the aligned
grains responsible for the starlight polarization.  Table
\ref{tab:results} gives the values of $\aeffp$ for each shape.

Figure \ref{fig:PhiPSA}a shows $\Phi_\PSA$ for each of the six cases
studied.  \citet{Draine_2024b} found that the parameter
\beq
\psi\equiv(\Asymm-1)^{0.6}(1-\poromacro)^{0.8}
\eeq
is a good predictor of $\Phi_\PSA(\aeffp)$.  The prolate and oblate shapes
fall close to the trendlines
\beqa \label{eq:oblate trend}
\Phi_\PSA &\,\approx\,\,& 1.35(\psi-0.07)
\hspace*{1.0cm}{\rm for~} \Stretch<1.5 ~~{\rm (oblate)}
\\ \label{eq:prolate trend}
\Phi_\PSA &\approx& 1.0(\psi-0.07)
\hspace*{1.2cm}{\rm for~} \Stretch>1.5 ~~{\rm (prolate)}
~~~.
\eeqa
The spheroidal analogs using the Bruggeman EMT have polarization
cross sections which come close to reproducing $\Phi_\PSA$ calculated
directly for the random aggregates.  We see in Figure
\ref{fig:PhiPSA}a that approximating the aggregates by spheroids using
the Bruggeman EMT results in reasonably accurate estimates of
$\Phi_\PSA$ for both flattened and elongated irregular aggregates.

To reproduce the observed strength of starlight polarization, the
grains with $\aeff\gtsim 0.10\micron$ must have $\Phi_\PSA\gtsim 0.7$
\citep{Draine+Hensley_2021c}; three of the six cases (\#2, \#3, \#5)
meet that condition (see Figure \ref{fig:PhiPSA}).

\subsection{Starlight Polarization Width Parameter $\sigmap$}

The individual grains producing the polarization must have
sufficiently narrow polarization profiles so that the population of
aligned grains (with $0.1\micron\ltsim\aeff\ltsim 0.3\micron$) can
produce a profile matching observations \citep{Serkowski_1973}.  For
an individual grain, the width of the polarization profile can be
characterized by the dimensionless width parameter $\sigmap(\aeff,{\rm
  shape})$ defined by \citet[][Eq.\ 44]{Draine+Hensley_2021c}.  The
observed starlight polarization requires that the grains responsible
for the polarization have $\sigmap\ltsim 0.6$ \citep{Draine_2024a}.

Figure \ref{fig:PhiPSA}b shows $\sigmap$ evaluated for the 12
aggregates and their spheroidal analogs.  The four EMTs give nearly
identical results for $\sigmap$, generally tending to overestimate
$\sigmap$.

With $\sim$10\% accuracy, $\Phi_\PSA$ and $\sigmap$ provided by the
SAM are useful for testing candidate shapes for modeling starlight
polarization.

  
\subsection{Far-Infrared Absorption}

\begin{figure}
\begin{center}  
\includegraphics[angle=0,width=\fwidth,
                 clip=true,trim=0.5cm 5.0cm 0.1cm 2.5cm]
{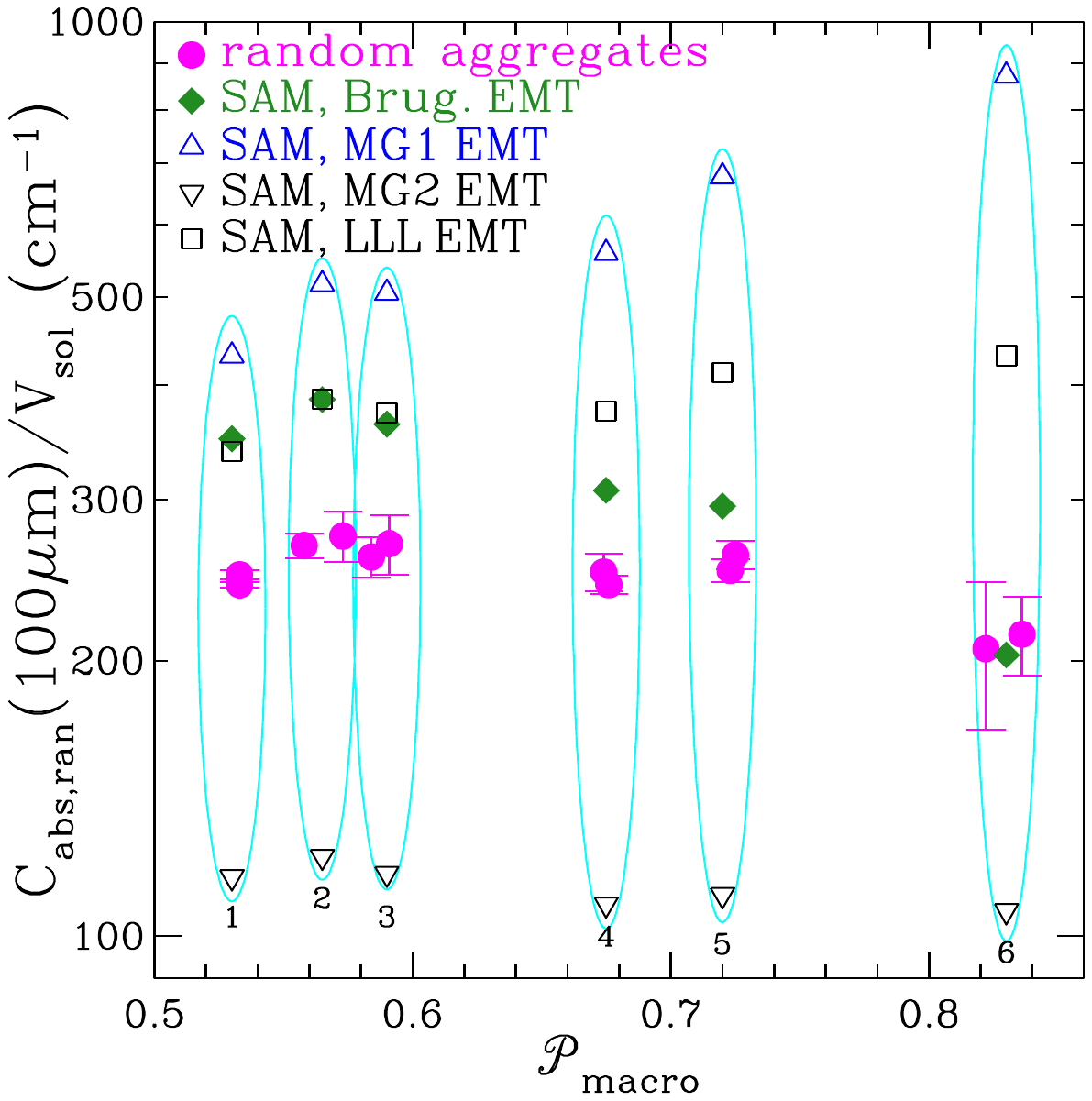}
\includegraphics[angle=0,width=\fwidth,
                 clip=true,trim=0.5cm 5.0cm 0.1cm 2.5cm]
{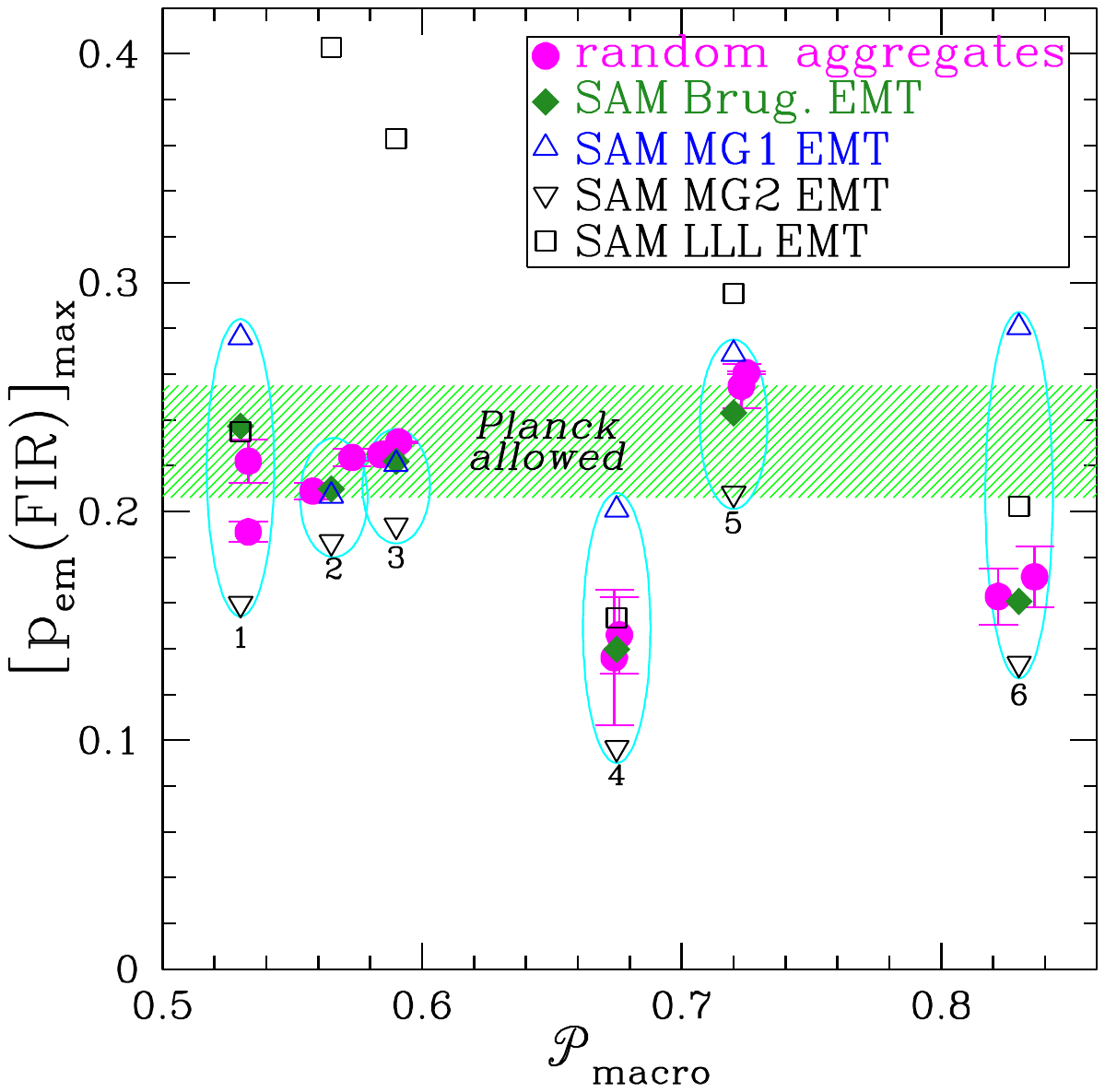}
\caption{\label{fig:Cabs/V}\label{fig:pfirmax}\footnotesize 
  {\bf (a)} Absorption cross section per unit solid volume for
  randomly-oriented grains at $\lambda=100\micron$, for twelve random
  aggregates (filled circles) versus $\poromacro$, for the six cases
  in Table \ref{tab:targets}.  Error bars show uncertainties in the
  DDA calculations.  Also shown are SAM results for $C_{\rm
    abs,ran}/V_\solid$, using different EMTs.  MG1 overestimates the
  absorption and MG2 underestimates the absorption.  The Bruggeman EMT
  is most accurate: the fractional error of $\sim$43\% for
  $\poromacro=0.53$ (case 1) decreases with increasing $\poromacro$;
  with excellent agreement for $\poromacro=0.83$ (case 6).
  {\bf (b)} Far-infrared and submm polarization, for fractional
  alignment given by Eq.\ (\ref{eq:falign}).  Results are shown for
  random aggregates in Table \ref{tab:targets}, grouped into six cases
  (labeled), together with spheroidal analogs.  The shaded area
  shows the allowed region $[p_{\rm em}({\rm FIR})]_{\rm max}$
  \citep{Planck_2018_XII}.
  }
\end{center}
\end{figure}

For submicron grains at $\lambda\gtsim 10\micron$, scattering is
negligible and $C_{\rm abs} \approx C_{\rm ext}$.  At wavelengths
$\lambda\gtsim10\micron$, where $|\epsilon-1|$ becomes large, the
different EMTs (see Figure \ref{fig:eps_emt}) provide very different
estimates for $\epsilon_{\rm EMT}(\lambda)$.  Not surprisingly, the
calculated absorption cross sections for the spheroidal analogs also
differ.  This was evident in Figures \ref{fig:Q_emt0.53}a and
\ref{fig:Q_emt0.83}a, where, for $\lambda\gtsim10\micron$, $\lambda
Q_\extran/\aeff$ differs widely for the different EMTs.

Figure \ref{fig:Cabs/V} shows $C_{\rm abs,ran}(100\micron)/V$ for the
six cases, plotted against $\poromacro$.  The Bruggeman EMT generally
provides the best results, but even the Bruggeman EMT overestimates
the absorption by more than 40\% for the two cases with $\poromacro <
0.58$.  For increasing $\poromacro$, the Bruggeman EMT works well; the
LLL EMT continues to overestimate the absorption, while the two
versions of Maxwell Garnett underestimate (for MG2) and overestimate
(for MG1) the absorption by large factors.

It is striking that $C_{\rm abs,ran}(100\micron)/V_\solid$ calculated
directly for the aggregates is nearly independent of $\poromacro$ over
the range 0.53--0.83, with $C_{\rm
  abs,ran}(100\micron)/V_\solid\approx 240\pm30 \cm^{-1}$ for the 12
aggregates studied.  A single isolated sphere with the dielectric
function assumed for the solid material in the aggregates would have
\beq
\frac{\Cabs(100\micron)}{V_\solid} =
\frac{18\pi}{\lambda} 
\frac{{\rm Im}(\epsilon_{\rm Ad})}
     {({\rm Re}(\epsilon_{\rm Ad})+2)^2 + ({\rm Im}(\epsilon_{\rm Ad}))^2}
\approx 104\cm^{-1} 
~~~.
\eeq
Aggregation raises the far-infrared (FIR) opacity by a factor
$\sim$$2.3$, but the overall porosity $\poromacro$ seems to have
little effect on the absorption for random orientations.

\subsection{Far-Infrared Polarization}

\citet{Planck_2018_XII} showed that the observed thermal emission from
interstellar dust reached polarization fractions as large as $[p_{\rm
    em}(850\micron)]_{\rm max}=0.220_{-0.014}^{+0.035}$.  The
fractional polarization is expected to be almost independent of
wavelength for FIR wavelengths, $\lambda \gtsim 50\micron$
\citep[see][Figure 11]{Draine_2024a}.  

Maximum polarization occurs when the magnetic field is perpendicular
to the line-of-sight.  If a mass-weighted fraction $\falign$ is
perfectly aligned, and the remainder randomly oriented, then
\beq
 [p_{\rm em}({\rm FIR})]_{\rm max}
=
\frac{\falign C_\polPSA({\rm FIR})}
{\falign C_\extPSA({\rm FIR}) + (1-\falign)C_\extran({\rm FIR})}
~~~,
\eeq
where $\CextPSA$ is the absorption cross section for perfect spinning
alignment.  The weakness of starlight polarization in the ultraviolet
requires that the small grains not be aligned. Because the small
nonaligned grains are estimated to contribute $\sim$30\% of the dust
mass, this implies $\falign\ltsim0.7$.  To reproduce the starlight
polarization, the aligned fraction is \citep{Draine+Hensley_2021c}
\beq
\falign \approx \frac{0.49}{\Phi_\PSA}
~~~.
\eeq
Therefore, we take
\beq \label{eq:falign}
\falign= {\rm min}\left(\frac{0.49}{\Phi_\PSA},0.7\right)
\eeq
to evaluate
$[p_{\rm em}({\rm FIR})]_{\rm max}$,
plotted in Figure \ref{fig:pfirmax}b.  For all six cases studied, the
values of
$[p_{\rm em}({\rm FIR})]$
calculated using Bruggeman EMT spheroids are in good agreement with
the random aggregates that the spheroids are intended to emulate.

The LLL, MG1, and MG2 spheroidal analogs generally show larger
discrepancies for
$[p_{\rm em}({\rm FIR})]_{\rm max}$
than the Bruggeman EMT-spheroid -- see Figure \ref{fig:pfirmax}b.

In Figure \ref{fig:pfirmax}b, three cases (\#2, \#3, and \#5) are
consistent with the Planck polarization.  Note that these are the same
cases that have $\Phi_\PSA > 0.7$ (see Figure \ref{fig:PhiPSA}), so
that they are also able to reproduce the polarization of starlight.

\begin{figure}
\begin{center}   
\includegraphics[angle=0,width=\fwidthc,
                 clip=true,trim=0.5cm 5.0cm 0.1cm 4.2cm]
{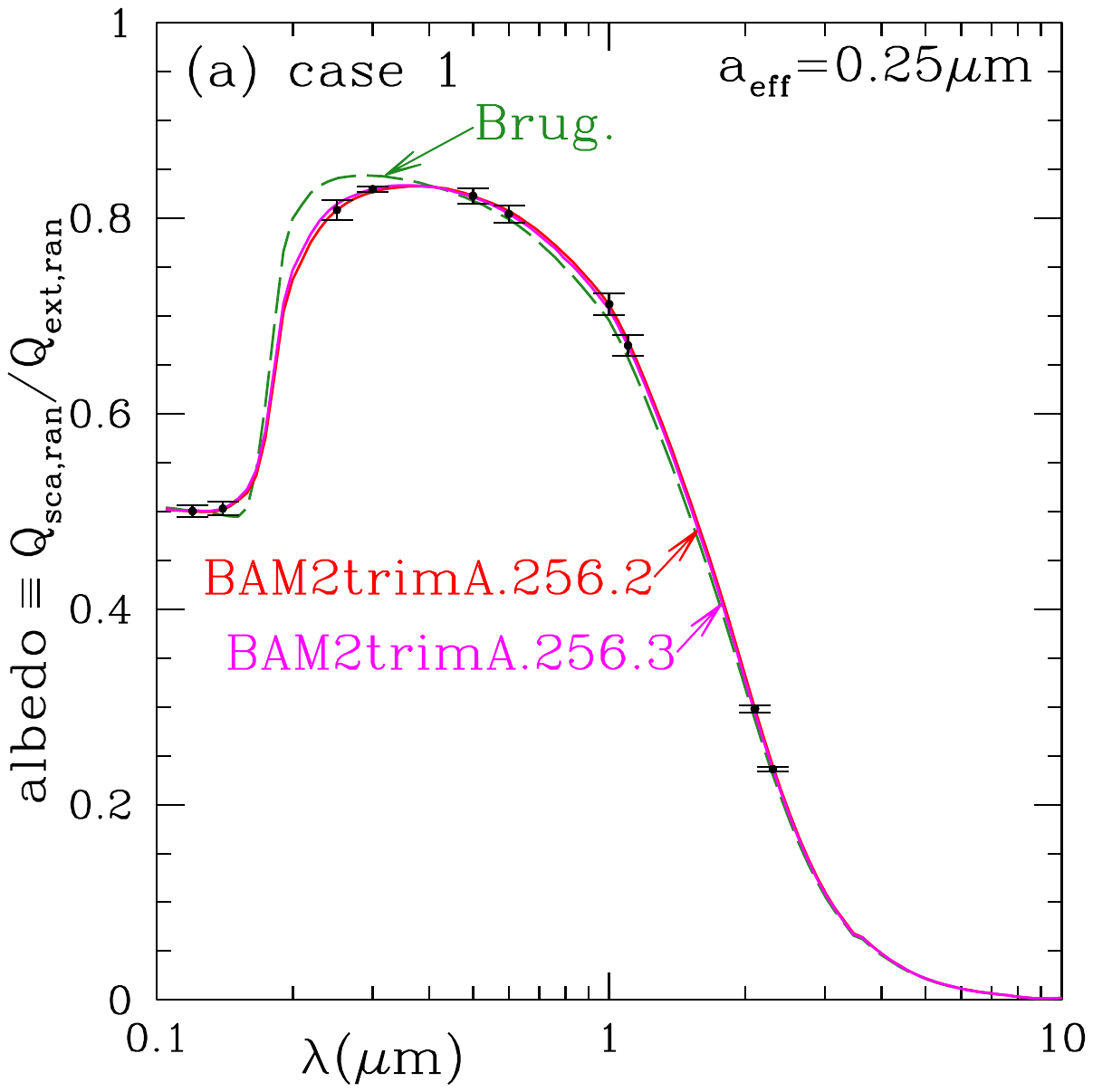}
\includegraphics[angle=0,width=\fwidthc,
                 clip=true,trim=0.5cm 5.0cm 0.1cm 4.2cm]
{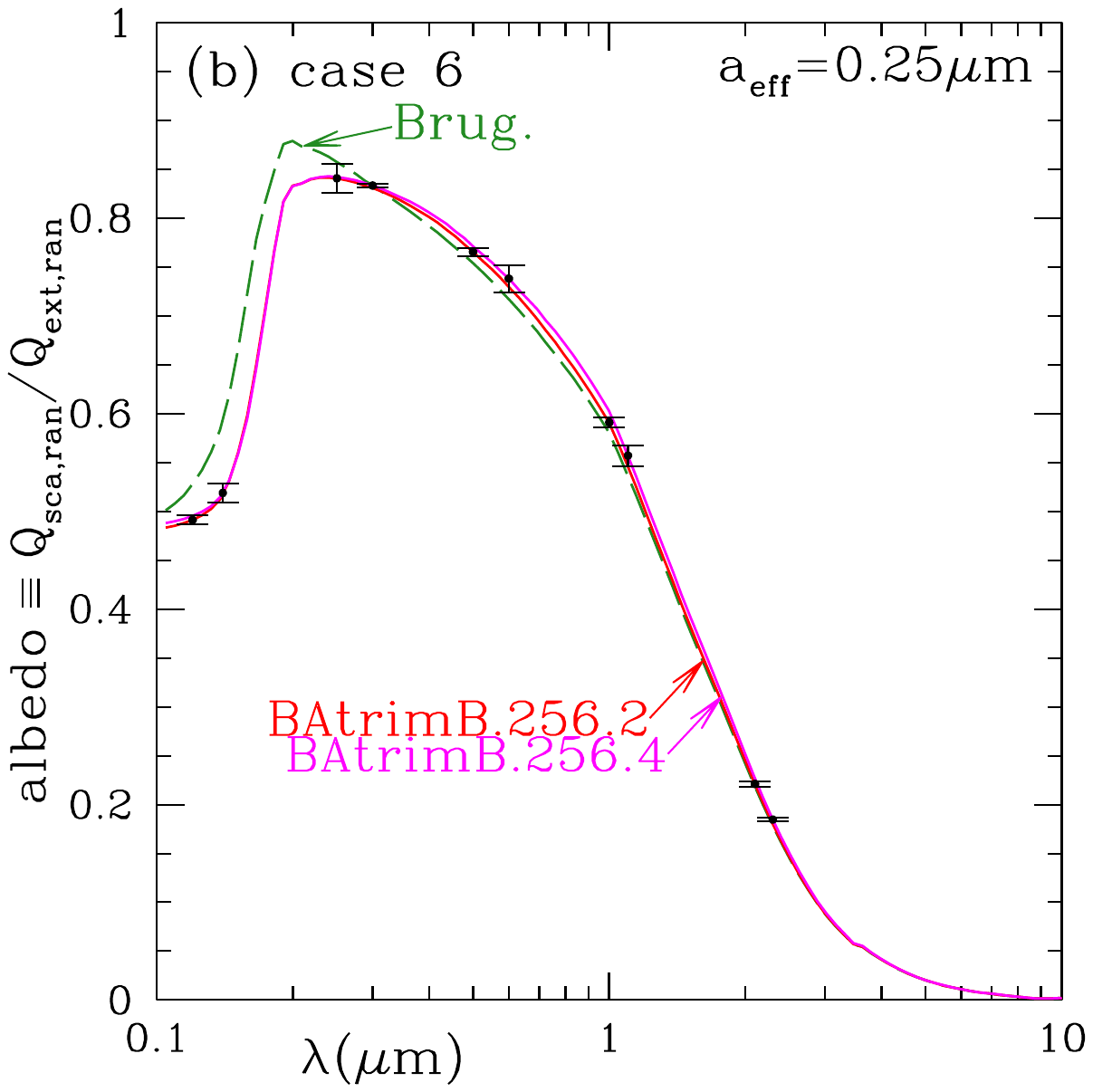}
\caption{\label{fig:albedo} Albedo for {\bf (a)} case 1 and {\bf (b)}
  case 6 aggregates, randomly oriented, with $\aeff=0.25\micron$.
  Solid curves: results for the aggregates; at selected wavelengths,
  error bars show the (small) uncertainties in DDA calculations for
  each aggregate.  Dashed curve: result for the SAM using the
  Bruggeman EMT.
  }
\end{center}
\end{figure}
\begin{figure}
\begin{center}  
\includegraphics[angle=0,width=\fwidthc,
                 clip=true,trim=0.5cm 5.0cm 0.1cm 4.2cm]
{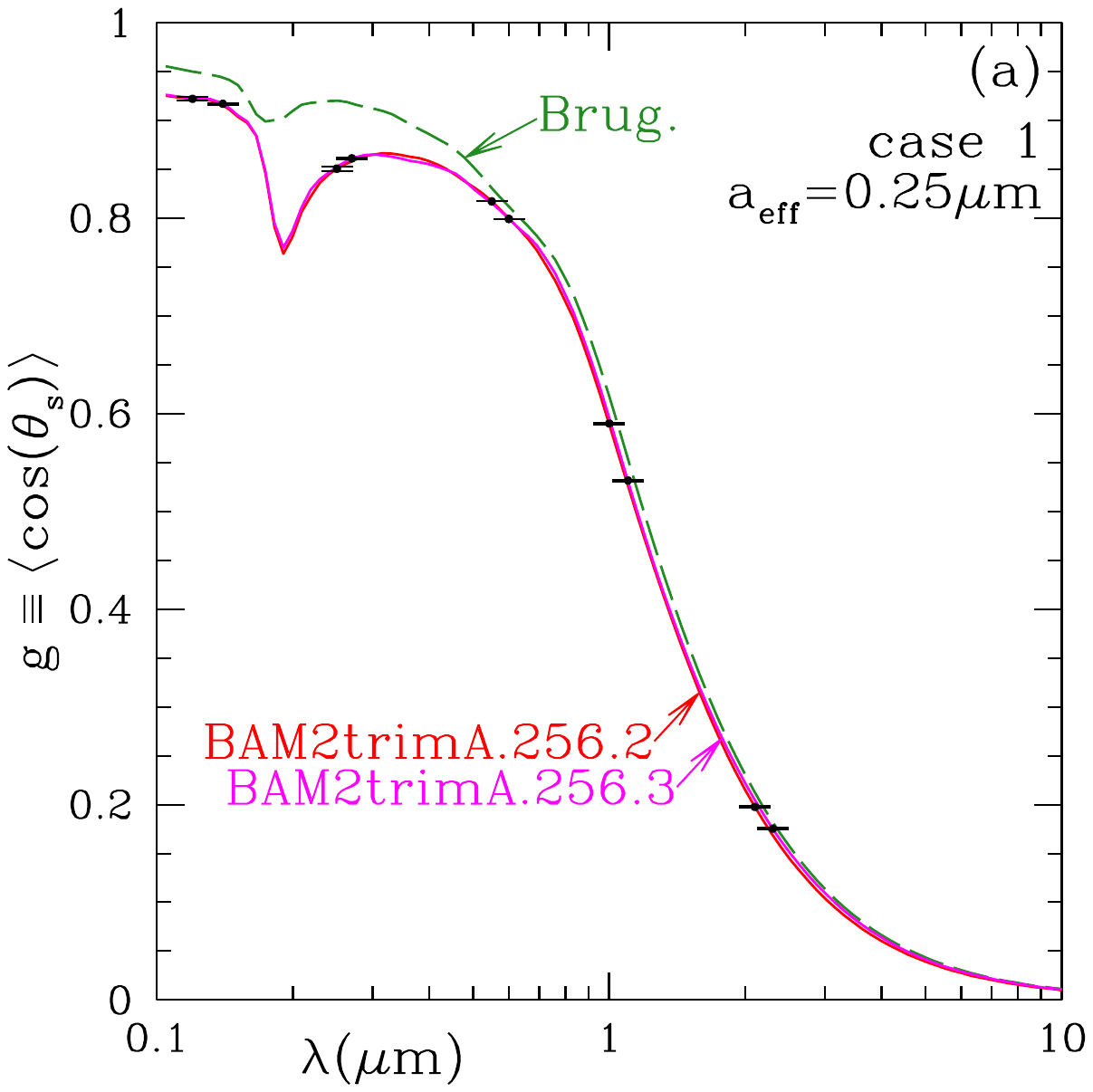}
\includegraphics[angle=0,width=\fwidthc,
                 clip=true,trim=0.5cm 5.0cm 0.1cm 4.2cm]
{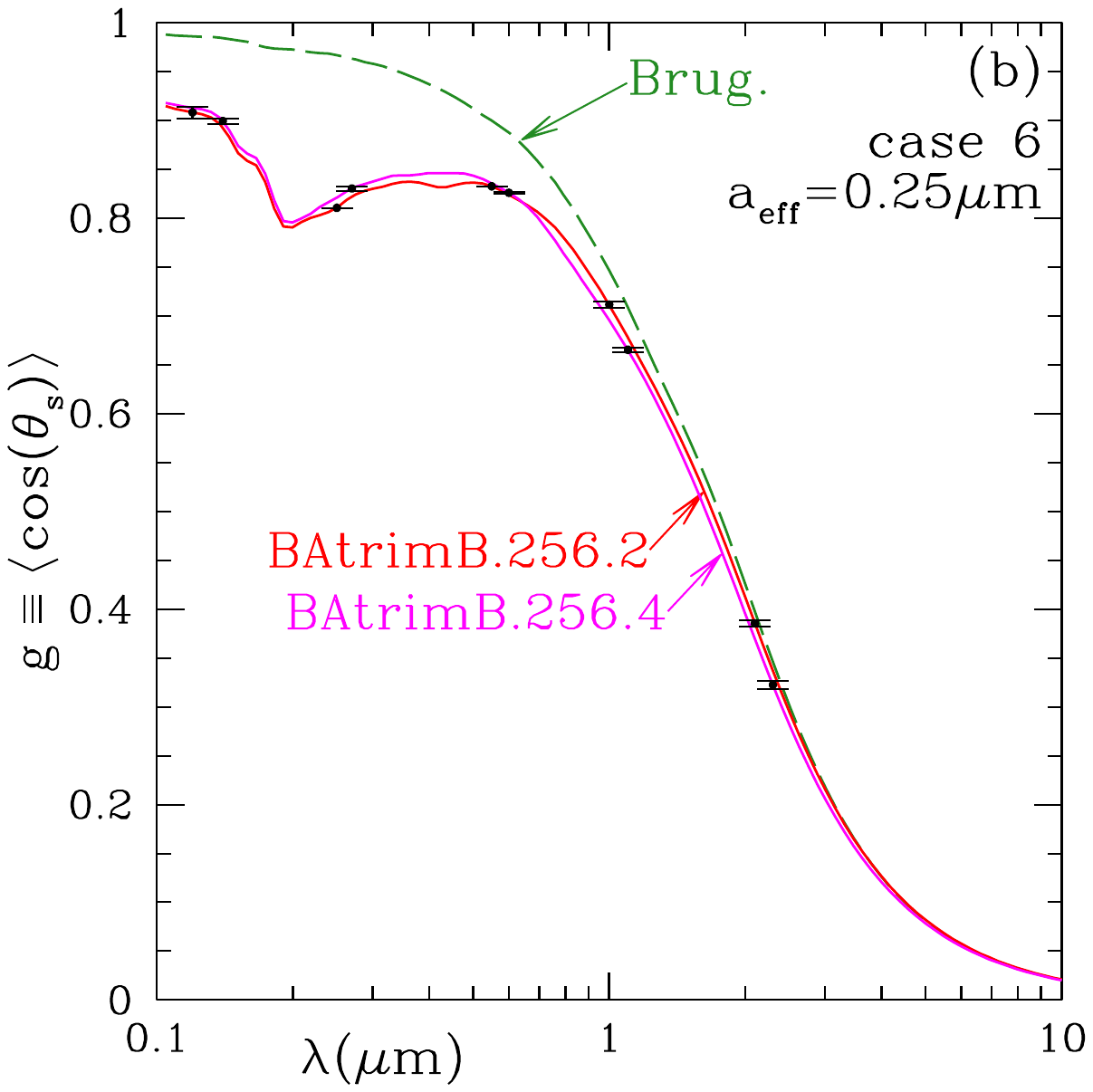}
\caption{\label{fig:<cos>} Same as Figure
  \ref{fig:albedo}, but for the scattering asymmetry factor
  $g=\langle\cos\thetas\rangle$.
  }
\end{center}
\end{figure}

\subsection{Scattering Properties}

The SAM with the Bruggeman EMT has been found to provide a useful
approximation to the extinction cross sections for the irregular
aggregate.  We now use the DDA to see how well it reproduces the
light-scattering properties of the aggregates.

Scattering depends on grain size; here we show results for
$\aeff=0.25\micron$.  Figure \ref{fig:albedo} shows the albedo
\beq
{\rm albedo}(\lambda) ~\equiv
\frac{\Csca(\lambda)}{\Cabs(\lambda)+\Csca(\lambda)}
\eeq
for randomly oriented aggregates in ``case 1'' ($\poromacro\approx
0.53$) and ``case 6'' ($\poromacro\approx0.83$).  For both cases the
albedo is $\sim$$0.5$ at $\lambda = 0.1\micron$ because the grain
material is strongly absorbing (see the rise in ${\rm Im}(\epsilon)$
at short wavelengths in Figure \ref{fig:eps_emt}).  The albedo rises
to $\sim$$0.8$ at $\lambda\approx 0.3\micron$, and declines at long
wavelengths as the grain (with $\aeff/\lambda \ll 1$) becomes
ineffective at scattering.  The SAM result for the albedo (using
the Bruggeman EMT) is in good agreement with the actual albedos of the
aggregates.

Figure \ref{fig:<cos>} shows the scattering asymmetry parameter
$g\equiv\langle\cos\thetas\rangle$ as a function of wavelength for
case 1 and case 6.  The SAM provides good overall agreement,
although tending to overestimate $g$ for shorter wavelengths.

\begin{figure}
\begin{center}  
\includegraphics[angle=0,width=\fwidthc,
                 clip=true,trim=0.5cm 0.5cm 0.1cm 2.0cm]
{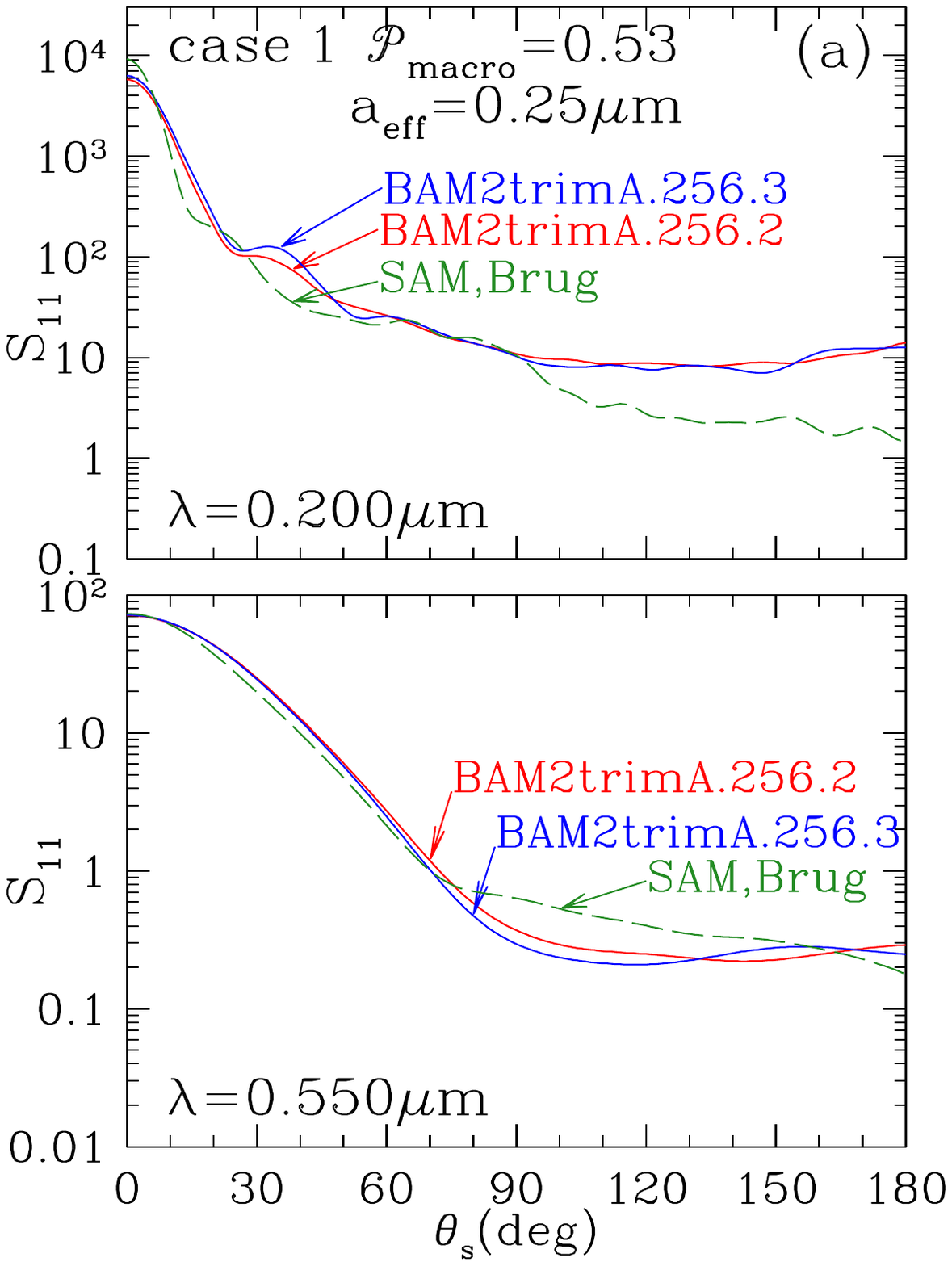}
\includegraphics[angle=0,width=\fwidthc,
                 clip=true,trim=0.5cm 0.5cm 0.1cm 2.0cm]
{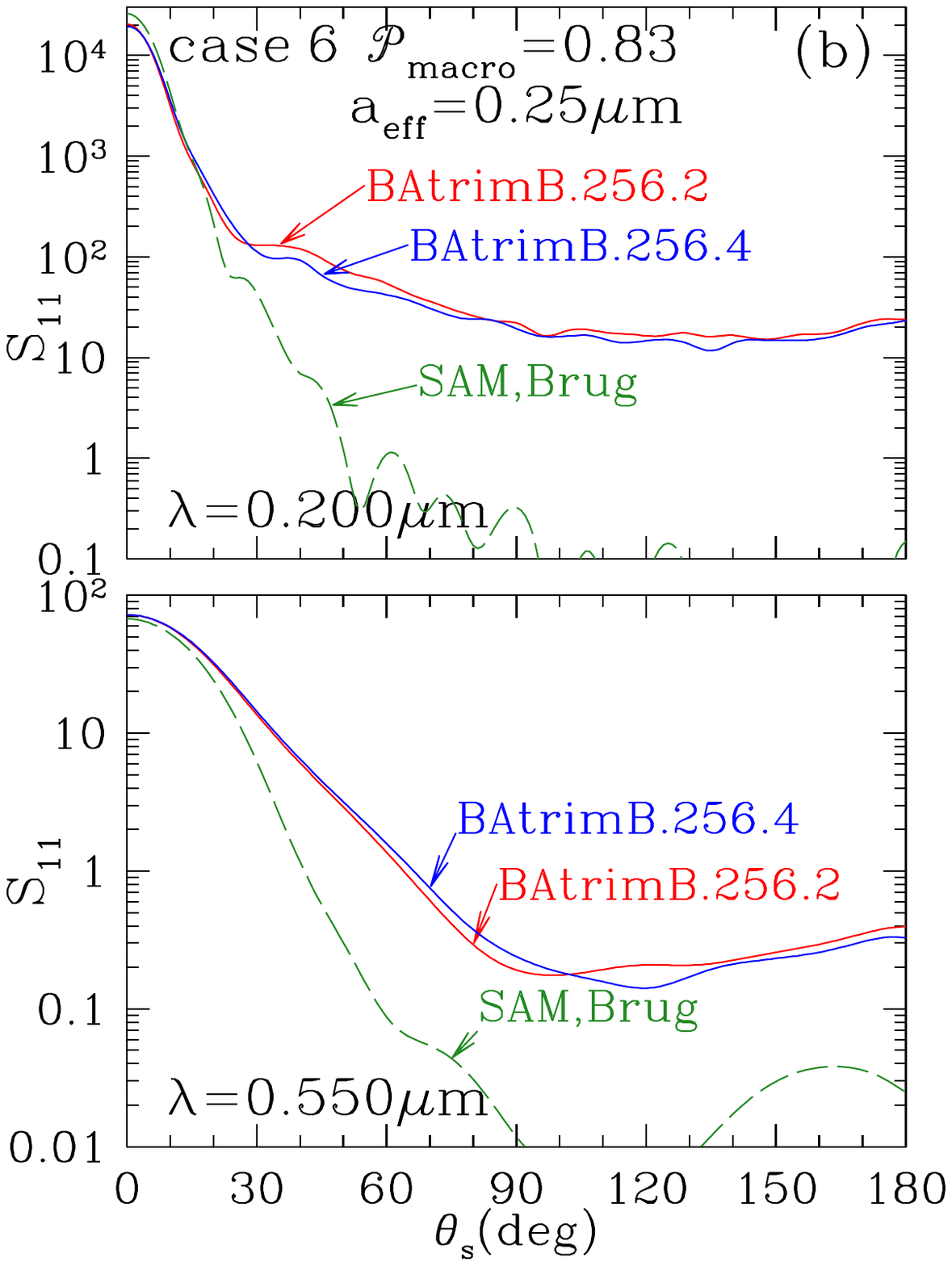}
\caption{\label{fig:S11} Scattering phase function $S_{11}$ as a
  function of scattering angle $\thetas$, for randomly-oriented
  aggregates with $\aeff=0.25\micron$, for $\lambda=0.2\micron$ and
  $0.55\micron$.  {\b (a)} Case 1 ($\poromacro=0.53$); {\bf (b)} Case
  6 ($\poromacro=0.83$).  Solid curves: aggregates; dashed curve:
  SAM using the Bruggeman EMT.
  }
\end{center}
\end{figure}

In addition to the albedo and scattering asymmetry
$\langle\cos\thetas\rangle$, it is of interest to compare the 
scattering phase functions for the random aggregates.  Figure
\ref{fig:S11} shows the Mueller scattering matrix element
\beq
S_{11}(\thetas)\equiv 
\frac{4\pi^2}{\lambda^2}\frac{dC_{\rm sca}(\thetas)}{d\Omega}
\eeq
calculated for randomly-oriented aggregates, where $dC_{\rm
  sca}(\thetas)/d\Omega$ is the differential scattering cross section
for scattering angle $\thetas$.  Figure \ref{fig:S11} shows $S_{11}$
for the case 1 and case 6 aggregates with $\aeff=0.25\micron$ for two
selected wavelengths: $\lambda=0.2\micron$, and $0.55\micron$.  For
$\lambda=0.2\micron$, the $\aeff=0.25\micron$ aggregates are very
forward-scattering.

Figure \ref{fig:S11} also shows $S_{11}$ calculated for the spheroidal
analogs, using the Bruggeman EMT.  For the case 1 aggregates,
$S_{11}$ for the spheroidal analog is in general agreement with the
aggregates for $\thetas\ltsim 90^\circ$ (most of the scattered power).
For the higher porosity case 6 aggregates, the SAM systematically
underestimates $S_{11}$ for $\thetas \gtsim \lambda/(2\aeff) =
63^\circ(\lambda/0.55\micron)(0.25\micron/\aeff)$.  Nevertheless, the
SAM yields fairly accurate estimates for the directions that dominate
the scattering, as already seen in the good agreement for the albedo
and scattering asymmetry factor $g=\langle\cos\thetas\rangle$ (see
Figures \ref{fig:albedo} and \ref{fig:<cos>}).

\medskip
\section{\label{sec:discuss}
         Discussion}

The present study demonstrates that the optical properties of
irregular, porous aggregates can be approximated by the ``spheroidal
analog method'' (SAM): the irregular aggregate, with solid volume
$V_\solid$, is modeled by a homogeneous spheroid.  The SAM prescribes
the volume $(4/3)\pi ab^2$ and axial ratio $a/b$ of the spheroid from
$V_\solid$ and simple geometric measures
($\poromacro,\Asymm,\Stretch$) of the aggregate's shape (see Appendix
\ref{app:geom}).  The dielectric function is obtained from the
Bruggeman EMT for porosity $\poromacro$.  The relatively simple
geometry of the spheroidal analog enables efficient computational
methods \citep[e.g.,][]{Voshchinnikov+Farafonov_1993} to be used to
calculate cross sections for scattering and absorption of
electromagnetic radiation.

Twelve irregular aggregate shapes were studied, with
$0.53<\poromacro<0.84$.  For $\aeff=0.25\micron$, the SAM extinction
cross sections are generally accurate to within $\sim$$10\%$ for
wavelengths $0.3\micron \ltsim \lambda \ltsim 10\micron$ (see Figure
\ref{fig:err}a).

The starlight polarization efficiency integral $\Phi_\PSA$ and
starlight polarization width parameter $\sigmap$ are calculated with
accuracies of $\sim$$10\%$ (see Figure \ref{fig:PhiPSA}),
permitting a quick test of whether a given irregular shape could
account for the observed polarization of starlight.

The scattering albedo and $g=\langle\cos\thetas\rangle$ are reproduced
quite well (see Figures \ref{fig:albedo} and \ref{fig:<cos>}), and the
full scattering phase function is reproduced well for the scattering
angles that account for most of the scattered power (see Figure
\ref{fig:S11}).  \citet{Mishchenko+Dlugach+Liu_2016} found that
monomer size $a_0$ affects the scattering phase function when
$x_0\equiv 2\pi a_0/\lambda \gtsim 0.5$.  The aggregates studied here
have $a_0=\aeff/256^{1/3}$, thus the finite monomer size must affect
the phase functions shown in Figure \ref{fig:S11}, with $x_0=0.45$ at
$\lambda=0.55\micron$, and $x_0=1.24$ at $\lambda=0.2\micron$.
  
The goal is to approximate {\bf\emph{a population of irregular
  aggregates}}; each aggregate will deviate from a simple spheroid on
many scales, including scales comparable to the overall size
$\aeff/(1-\poromacro)^{1/3}$.  An individual aggregate may have
features in the scattering phase function reflecting its particular
geometry, but many of these details will be suppressed when averaging
over the population.  Even for individual aggregates, the albedo and
scattering asymmetry parameter $g\equiv\langle\cos\thetas\rangle$ are
quite well approximated by the SAM (see Figures \ref{fig:albedo} and
\ref{fig:<cos>}).

At longer wavelengths $\lambda \gtsim 10\micron$, the dielectric
function of the material becomes large, and the SAM becomes less
accurate, with fractional errors in absorption opacity for some cases
as large as $\sim$40\%.  There is clearly room for improvement;
perhaps some new EMT can be developed that can provide better results
in the far-infrared.  Nevertheless, the SAM with the Bruggeman EMT
already provides useful accuracy for fractional polarization, even in
the far-infrared.

In conclusion, the SAM is useful for modeling interstellar grains,
even if the grains themselves have complex, irregular shapes resulting
from aggregation.  It eliminates the need to select specific aggregate
structures, allowing the modeler to focus on the important structural
parameters, $\poromacro$, $\Asymm$, and $\Stretch$.  Grain models
\citep[e.g.,][]{Draine+Hensley_2021a, Draine+Hensley_2021c,
  Hensley+Draine_2023} that assume spheroidal shapes with variable
porosity can be used to estimate grain masses, porosities
$\poromacro$, and effective axial ratios $a/b$, even if the actual
grains are irregular aggregates.

In this paper the SAM has been applied to irregular aggregates
composed of a single type of solid (characterized by the ``astrodust''
dielectric function).  However, Bruggeman's approach to EMT is in
principle extensible to multiple components \citep{Sihvola_1999}: for
$N$ solid components, each with volume filling factor $f_j$ and
dielectric function $\epsilon_j$, Equation (\ref{eq:Brugg})
generalizes to
\beq \label{eq:NBrug}
0 = \sum_{j=0}^N f_j
\left(\frac{\epsilon_j-\epsilon_\Brug}{\epsilon_j+2\epsilon_\Brug}\right)
~~~,
\eeq 
where $j=0$ corresponds to vacuum, with $f_0=\poromacro$ and
$\epsilon_0=1$.\footnote{%
  Equation (\ref{eq:NBrug}) generally has $N+1$ solutions for
  $\epsilon_\Brug$.  For the convention $E\propto e^{-i\omega t}$,
  solutions with ${\rm Im}(\epsilon_\Brug)<0$ are unphysical and can
  be rejected; for $N=1$ this leaves the solution
  (\ref{eq:epsilon_Brug}).  However, for $N>1$ more than one solution
  may fall in the upper half-plane, and other arguments will be needed
  to decide which to use for $\epsilon_\EMT$.}
Thus the SAM could, in principle, be used to study aggregates composed
of more than one material, e.g., amorphous silicate and hydrogenated
amorphous carbon.


\section{\label{sec:summary}
         Summary}

The principal results are as follows:
\begin{enumerate}

\item The spheroidal analog method (SAM) presented here allows the
  optical properties of grains with complex geometries, including
  irregular aggregates, to be approximated -- with acceptable accuracy
  -- by the optical properties of simple spheroids with prescribed
  size, axial ratio, and effective dielectric function.

\item For dielectric functions relevant to interstellar grain
  materials, the SAM, with the Bruggeman EMT, provides good accuracy
  -- typically $\sim$10\% -- for wavelengths $\lambda \ltsim
  10\micron$.

\item At wavelengths $\lambda\gtsim20\micron$, where the dielectric
  function of interstellar grain material becomes large, the accuracy
  of the SAM degrades.  Absorption cross sections may be
  overestimated by up to $\sim$40\%, depending on the porosity of the
  irregular aggregate.  However, calculated fractional polarizations
  are more accurate.

\item For $\lambda < 10\micron$, where light scattering can be
  important for interstellar grains, the SAM provides accurate
  calculations of albedos and $g=\langle\cos\thetas\rangle$, and
  provides a good approximation to the scattering phase function for
  scattering angles that account for most of the scattered power.

\end{enumerate}
\begin{acknowledgements}

I am grateful to Dina Gutkowicz-Krusin for many helpful suggestions,
and to the anonymous referee for a thoughtful and helpful review.  I
thank Robert Lupton for availability of the SM package.

\end{acknowledgements}

\begin{appendix}
\section{\label{app:geom}
         $\poromacro$, $\Asymm$, and $\Stretch$}

Consider a grain of mass $M$ with arbitrary geometry, with effective
radius $\aeff\equiv(3M/4\pi\rho_\solid)^{1/3}$.  Let $I_1\geq I_2 \geq
I_3$ be the eigenvalues of the moment of inertia tensor, and define
\beq
 \alpha_j \equiv \frac{I_j}{0.4 M\aeff^2}
~~~.
\eeq
The geometric shape of the aggregate is characterized by the
``macroporosity'' \citep{Shen+Draine+Johnson_2008}
\beq
\poromacro
\equiv 1 -
\frac{1}{\left[(\alpha_2+\alpha_3-\alpha_1)
         (\alpha_1+\alpha_3-\alpha_2)(\alpha_1+\alpha_2-\alpha_3)\right]^{1/2}}
~~~,
\eeq
the ``asymmetry parameter'' \citep{Draine_2024b},
\beq
\Asymm \equiv
\left(\frac{\alpha_1}{\alpha_2+\alpha_3-\alpha_1}\right)^{1/2}
~~~,
\eeq
and the ``stretch parameter'' \citep{Draine_2024a}
\beq
\Stretch \equiv \frac{\alpha_2}{(\alpha_1\alpha_3)^{1/2}}
~~~.
\eeq

\section{\label{app:DDA}
         Using the Discrete Dipole Approximation}

The DDA calculations were carried out using the open source
code {\tt DDSCAT}\footnote{%
Version 7.3.3, available at \url{ddscat.wikidot.com}.}
When using {\tt DDSCAT} to directly calculate the scattering
properties of irregular aggregates (using, e.g., shape option {\tt
  SPHERES\_N}), the variable {\tt AEFF} $=\aeff$ given by Equation
(\ref{eq:aeff}).

When using DDSCAT to calculate scattering for the spheroidal analog:
\begin{enumerate}
\item Shape option {\tt ELLIPSOID} is used.

\item The effective radius parameter {\tt AEFF} $=
  \aeff/(1-\poromacro)^{1/3}$: the spheroidal analog has a larger
  volume than the volume $V_\solid$ of solid material in the
  aggregate.

\item Extinction, absorption and scattering efficiency
  factors {\tt Qext}, {\tt Qabs}, and {\tt Qsca} returned by DDSCAT
  need to be rescaled, e.g.:
\beq Q_\ext = {\tt Qext}\times(1-\poromacro)^{-2/3}
~~~,
\eeq
in order to be able to compare with $Q_\ext$ values calculated
directly for the aggregate (similarly for $Q_{\rm sca}$ and $Q_{\rm
  abs}$).

\end{enumerate}
For the results shown here, we considered 151 wavelengths, uniformly
spaced in $\log(\lambda)$ from $0.1\micron$ to $100\micron$.

For the irregular aggregates we used 11 values of the angle
$\Theta\in[0,\pi]$ (where $\Theta$ is the angle between the principal
axis $\bahat_1$ and the direction of propagation $\bkhat$), and 12
rotations of the aggregate around $\bahat_1$, for a total of 132
distinct orientations relative to the incident radiation.

For the spheroids, we used 11 values of the angle $\Theta\in[0,\pi/2]$
between the symmetry axis $\bahat$ and $\bkhat$.

Orientational averages were calculated for both random orientation and
perfect spinning alignment (PSA) with $\bahat_1\perp\bkhat$ as
described in Appendix D of \citet{Draine_2024a}.

\end{appendix}

\bibliography{/u/draine/work/libe/btdrefs}

\end{document}